\makeatletter
\declare@file@substitution{revtex4-1.cls}{revtex4-2.cls}
\makeatother

\documentclass[twocolumn]{aastex631}
\usepackage{xcolor}
\usepackage{amsmath}
\usepackage[mathscr]{euscript}
\usepackage{multirow}

\received{}
\revised{}
\accepted{}

\submitjournal{The Astrophysical Journal}

\shorttitle{Flow Dynamics and Morphology of FR-I Jets}
\shortauthors{Bhattacharjee et al.}

\begin{document}

\title{A Simulation Study of Low-Power Relativistic Jets: Flow Dynamics and Radio Morphology of FR-I Jets}
\author[0000-0002-2878-4025]{Ayan Bhattacharjee}
\affiliation{Department of Physics, College of Natural Sciences, UNIST, Ulsan 44919, Korea}
\author[0000-0002-5550-8667]{Jeongbhin Seo}
\affiliation{Department of Physics, College of Natural Sciences, UNIST, Ulsan 44919, Korea}
\affiliation{Los Alamos National Laboratory, Theoretical Division, Los Alamos, NM 87545, USA}
\author[0000-0002-5455-2957]{Dongsu Ryu}
\affiliation{Department of Physics, College of Natural Sciences, UNIST, Ulsan 44919, Korea}
\author[0000-0002-4674-5687]{Hyesung Kang}
\affiliation{Department of Earth Sciences, Pusan National University, Busan 46241, Korea}
\correspondingauthor{Hyesung Kang}\email{hskang@pusan.ac.kr}
\correspondingauthor{Dongsu Ryu}\email{dsryu@unist.ac.kr}

\begin{abstract}


Radio galaxies are classified into two primary categories based on their morphology: center-brightened FR-I and edge-brightened FR-II. It is believed that the jet power and interactions with the ambient medium govern the deceleration and decollimation of the jet-spine flows, which, in turn, influence this dichotomy. Using high-resolution, three-dimensional relativistic hydrodynamic simulations, we follow the development of flow structures on sub-kpc to kpc scales in kinetically dominant low-power relativistic jets. We find that the bulk Lorentz factor of the jet spine and the advance speed of the jet head, which depend on the energy injection flux and the jet-to-background density contrast, primarily determine the dynamics and structures of the jet-induced flows. The entrainment of ambient gas and the background density and pressure gradient may also play significant roles. To emulate radio morphology, we produce the synthetic maps of the synchrotron surface brightness for the simulated jets, by employing simple models for magnetic field distribution and nonthermal electron population and considering relativistic beaming effects at different inclination angles. Both the flow structures and radio maps capture the longitudinal and transverse structures of the jet-spine and shear layer, consistent with observations. We also compare different background effects and argue that the loss of pressure confinement beyond the galactic core may be a key factor in the flaring and disruption of FR-I jets. Our results confirm that mildly relativistic jets could explain the one-sidedness or asymmetries with the boosted main jet and deboosted counterjet pairs.

\end{abstract}


\section{Introduction}\label{s1}

Relativistic jets emitted from active galactic nuclei (AGNs) travel from very close to the central black hole beyond the galactic core region up to megaparsec distance. Along their way through the interstellar medium (ISM) of their host galaxies and then through the intracluster medium (ICM), the jets interact with the background medium. 
They emit radiation from different processes, observed at multiple wavelengths, ranging from radio waves to $\gamma$-rays \citep[see, e.g.,][for reviews]{begelman1984,fabian2012,marti2019,perucho2019b,blandford2019,hardcastle2020}. 

\citet{fanaroff1974} categorized radio jets based on the relative position of ``hot spots." The classification was determined by the ratio of the distance between the brightest regions on the opposite sides of the core to the total source extent. A ratio below 0.5 classified the source as FR-I, while a ratio above 0.5 indicated an FR-II source. For conciseness, we will refer to this ratio as the ``FR ratio".
Additionally, it was observed that sources with a radio luminosity below $L_{178}\approx 2.0\times 10^{25}$ W/Hz/sr at 178 MHz are typically FR-I (e.g., Centaurus A and Virgo A), while those with higher luminosity are predominantly FR-II (e.g., Cygnus A).

{
Analyzing observational data, \citet{godfrey2013} found a relationship between the kinetic power of the jet, $Q_j$, and the radio luminosity of FR galaxies, $L_R$, expressed as $Q_j\propto L_R^{\alpha}$, where $\alpha \approx 0.64-0.67$. Moreover, their Figure 3 suggests that the divide between FR-I and FR-II occurs at approximately $Q_{\rm div}\sim 10^{45}-10^{46} {\rm erg~s^{-1}}$.
On the other hand, \citet{ledlow1996} found that the radio luminosity at the FR-I/II divide, $L_{\rm R,div}$, is correlated with the optical luminosity of the host galaxy, $L_{\rm opt}$, following the relation $L_{\rm R,div} \propto L_{\rm opt}^2$. This finding implies that the properties of the host galaxy may also influence the FR morphology.
More recently, \citet{mingo2019} demonstrated that the FR morphological divide cannot be solely described by $L_R$, as there is a significant overlap between FR-I and FR-II types in the $L_R-L_{\rm opt}$ plane (see their Figure 11).
These works along with many previous studies collectively establish that both the jet power $Q_j$ and the ambient medium play crucial roles in determining the jet morphology \citep[see][and references therein]{hardcastle2020}.}

Typically, the dynamical structures of FR-I jets consist of three distinct regions \citep{laing2002a,laing2002b}: {the inner jet-spine region characterized by a slow expansion of width, the cocoon region which undergoes a rapid change in width}, and the outer region that smoothly expands in a conical shape. These jets show stratification of velocity and brightness along the transverse direction, consistent with a spine/sheath structure \citep{laing2014}. Furthermore, some FR-I jets exhibit lobed structures at the far ends and asymmetric bright regions near the core \citep{laing1999}, both of which can be explained by the relativistic Doppler beaming of a pair of decelerating relativistic jets viewed at an inclined angle \citep{laing1999,laing2011}.

In general, FR-II jets, which have higher powers and hence higher luminosities, appear well collimated, maintain relativistic speeds, and terminate at hot spots.
On the other hand, FR-I jets with lower powers and lower luminosities tend to be decollimated and decelerate to subrelativistic speeds as they propagate out to kiloparsec scales \citep[e.g.,][]{bicknell1995,laing2014,mingo2019}.
Such dynamical differences between FR-I and FR-II jets are thought to be primarily governed by the jet power, $Q_j$ (see Eq. [\ref{Qj}]) \citep[e.g.,][]{kaiser1997,perucho2007,godfrey2013} and the momentum injection rate, $\dot{M}_j$, (see Eq. [\ref{Mjet}]) \citep[e.g.,][]{perucho2014,hardcastle2020}.
In other words, low-power FR-I jets with smaller $Q_j$ and $\dot{M}_j$ penetrate the ambient gas more slowly, resulting in more significant deceleration and decollimation.

The entrainment of the dense ambient gas is expected to expedite the deceleration of the jet flow in FR-I jets \citep[e.g.,][]{perucho2007,perucho2014}. In relativistic jets, mixing layers are thought to form between the relativistic, forward-moving jet-spine flow and the subrelativistic, backward-moving backflow through Kelvin-Helmholtz instability \citep[e.g.,][]{perucho2005,perucho2010}, Rayleigh-Taylor instability \citep[e.g.,][]{matsumoto2013,matsumoto2017}, and relativistic centrifugal instability \citep[e.g.,][]{gourgouliatos2018a,gourgouliatos2018b}.
Through turbulent mixing processes, some of the ambient gas is loaded onto the jet flow, leading to the deceleration of the jet. 
{Such dynamical evolution of FR-I jets has been extensively studied through relativistic hydrodynamic (RHD) simulations \citep[e.g.][]{perucho2007,rossi2008,english2016,li2018} and relativistic magnetohydrodynamic (RMHD) simulations \citep[e.g.][]{leismann2005,porth2015,marti2016,tchekhovskoy2016,massaglia2022,rossi2024}.}

For low-power FR-I jets, the mass loading caused by the injection of ionized plasma from stellar winds within the host galaxy could contribute to additional entrainment and deceleration \citep[e.g.,][]{komissarov1994, bowman1996}. { This effect is expected to be most pronounced within the galactic core, where early deceleration occurs due to the dissipation of kinetic energy \citep[e.g.,][]{perucho2014,angles2021}.
Additionally, \citet{perucho2020} proposed an alternative model for FR-I jet deceleration, suggesting that stars crossing the jet-ambient boundary could induce instabilities, which gradually grow and disrupt the jets.}

Furthermore, the loss of the confinement due to the background, after a jet moves into a surrounding halo with a declining density/pressure gradient, can cause the jet-induced flow to expand, which can lead to the deceleration and flaring \citep{begelman1984}. 
This phenomenon of breakout and expansion of the jet under changes of the background medium has been studied, using numerical simulations \citep[e.g.,][]{hooda1996, meliani2008,tchekhovskoy2016, mandal2022,perucho2023}. 
 
In numerical simulations, the dimensionality significantly influences the flow dynamics of relativistic jets, owing to inherent limitations in one and two dimensions. In \citet{bodo1998}, it was shown that three-dimensional (3D) jets exhibits more rapid and efficient mixing compared to their two-dimensional (2D) counterparts. Additionally, the entrained mass in 2D jets scales linearly with radial extent, whereas in 3D jets, it scales quadratically. Moreover, \citet{massaglia2016} demonstrated that under the same jet parameters, 2D simulations lead to the formation of more recollimation shocks in predominantly collimated jets, whereas 3D simulations result in fully turbulent jets with more significant deceleration. 

Recently, we developed the HOW-RHD code\footnote{HOW-RHD stands for High-Order WENO-based Relativistic HydroDynamic.}, a new 3D code that incorporates the fifth-order accurate, finite-difference WENO (weighted essentially non-oscillatory) scheme for solving hyperbolic conservation equations and the fourth-order accurate SSPRK (strong stability-preserving Runge-Kutta) scheme for time advance, along with the equation of state (EOS) that can closely approximate the EOS of the perfect gas in the relativistic regime \citep [see][for details]{seo2021a}. Using the code, \citet[][hereafter Paper I]{seo2021b} carried out 3D RHD simulations of high-power FR-II type jets and studied flow dynamics, such as shocks, turbulence, and velocity shear, produced in the jet-induced flows. 

In this paper, using the same HOW-RHD code, we perform 3D simulations of relativistic jets, focusing on low-power FR-I types. Additionally, we include mass loading from stellar winds in one of the jet models and consider different background density profiles. We analyze the structures and dynamics of the jet-induced flows as well as their deceleration and decollimation. Moreover, by adopting simple but physically motivated prescriptions for magnetic field distribution and nonthermal electron population, we estimate the synchrotron emissivity of the jet-induced flows. We then generate synthetic radio maps of the simulated jets viewed at different inclination angles.

The paper is organized as follows. In Section \ref{s2}, the details of simulation setups and model parameters are specified. Section \ref{s3} describes the simulation results, including the structures and dynamics of jet-induced flows and the morphological characteristics of synthetic radio maps.
Finally, in Section \ref{s4}, we provide a brief summary.

\section{Numerical Simulations}\label{s2}

\subsection{Basic Equations}\label{s2.1}

The RHD equations describing the jet development with mass loading can be written as follows
\citep[e.g.,][]{landau1959}:
\begin{eqnarray}
\frac{\partial D}{\partial t} + \mbox{\boldmath$\nabla$}\cdot(D\mbox{\boldmath$v$}) = S_D, \label{Deq}\\
\frac{\partial \mbox{\boldmath$M$}}{\partial t} + \mbox{\boldmath$\nabla$}\cdot(\mbox{\boldmath$M$}\mbox{\boldmath$v$}+p) = {\mbox{\boldmath$S$}}_M, \label{Meq}\\
\frac{\partial E}{\partial t} + \mbox{\boldmath$\nabla$}\cdot[(E+p)\mbox{\boldmath$v$}] = S_E\label{Eeq},
\end{eqnarray} 
where $D=\Gamma\rho$, $\mbox{\boldmath$M$}=\Gamma^{2}\rho(h/c^2) \mbox{\boldmath$v$}$, and $E=\Gamma^{2}\rho h -p$ are the mass, momentum, and total energy densities in the computational frame, respectively. 
Here, $c$ is the speed of light, $\Gamma=1/\sqrt{1-(v/c)^{2}}$ is the Lorentz factor, $v$ is the flow speed, and $h = (e + p)/\rho$ is the specific enthalpy, with $e=\epsilon+\rho c^2$ being the sum of the internal energy density, $\epsilon$, and the rest-mass energy density.  
{Since we consider a single-species baryonic fluid, we assume that both the jet inflow and the ambient gas consist of electrons and protons.}

The source terms on the right-hand side incorporate mass loading from stellar winds, as well as the effects of background stratification. Assuming that the loaded mass possesses negligible momentum and internal energy, as compared to the jet, it contributes only to the mass and rest-mass energy terms. The external gravity, which is required to balance the pressure gradient in the stratified background medium, contributes to the momentum and energy source terms. Combining these, the source terms are written as
\begin{eqnarray}
S_D = q_D,\label{Dsource}\\
{\mbox{\boldmath$S$}}_M = \rho \mbox{\boldmath$g$},\label{Msource}\\
S_E = q_D c^2 + \rho \mbox{\boldmath$v$}\cdot\mbox{\boldmath$g$},\label{Esource}
\end{eqnarray}
where $q_{\rm D}$ is the mass-loading rate per unit volume, and the external gravity, $\mbox{\boldmath$g$}=(\mbox{\boldmath$\nabla$}p_b)/\rho_b$, is set up with the initial density and pressure profiles of the background medium (see Section \ref{s2.2}).
{These source terms are applied to the entire computational box.}

As in Paper I, simulations are performed using the HOW-RHD code based on the WENO and SSPRK schemes. {The RC (Ryu-Chattopadhyay) version of the EOS} is employed to accurately follow the thermodynamics of relativistic fluids \citep{ryu2006}, along with several treatments to improve the accuracy and stability of the simulations.

\subsection{Background Stratification and Mass Loading}\label{s2.2}

As the fiducial model for stratified background media, we adopt the King profile of the density,
\begin{equation}
\rho_b(r) = \rho_c\left[ 1+ \left( \frac{r}{r_{c}}\right)^{2}\right]^{-3\beta_K/2},
\label{kings}
\end{equation}
where $r$ is the radial distance from the center of the host galaxy. 
We choose the setup of \citet{perucho2014}, which is designed to model the FR-I radio galaxy 3C 31 in the host galaxy, NGC 383 \citep{hardcastle2002}: $\beta_K=0.73$, $r_c=1.2$ kpc, and $\rho_c = 3.0\times10^{-25}~{\rm g~cm}^{-3}$. The background medium is assumed to be isothermal with $T_b=4.9\times10^6$ K. Assuming the background gas of fully ionized hydrogen, the pressure is then given as $p_b(r)=2\rho_b(r)k_BT_b/m_p$, where $k_{B}$ is the Boltzmann constant and $m_p$ is the proton mass: $p_c = 2\rho_c k_BT_b/m_p=2.4\times10^{-10}~{\rm dyn~cm}^{-2}$.

It is worth noting that in uniform background media, jets form scale-free systems, which can be scaled up and down to address jets at different length scales. However, in stratified media, jets are not scale free, and simulations depend on the radius of the jet at injection, $r_j$, and the injection position, $r_{\rm inj}$, from which the jet is launched. We consider two cases: $r_j=10$ pc and $r_{\rm inj}=80$ pc as the fiducial case \citep{perucho2014}, and $r_j=100$ pc and $r_{\rm inj}=250$ pc for larger scale FR-I jets, which are tracked beyond the galactic core. In the former, $\rho_b$ and $p_b$ at $r_{\rm inj}$ are $99.3\%$ of $\rho_c$ and $p_c$, while in the latter, they are $95.5\%$  of $\rho_c$ and $p_c$.

We also consider the background profile of a simple power-law distribution, specifically the one designed to model Centaurus A in the elliptical galaxy NGC 5128 \citep{wykes2019}: $\rho_b(r),~p_b(r) \propto \ r^{-3/2}$ with the density and pressure at $r_{\rm inj}$ matching those of the fiducial model, that is, $99.3\%$ of $\rho_c$ and $p_c$.

For the source term, $q_D(r)$, representing mass loading from stellar winds (denoted by the subscript `sw'), we adopt the prescription of \citet{perucho2014}:
\begin{equation}
q_D(r) = q_{\rm sw}\left[ 1+ \left( \frac{r}{r_{\rm sw}}\right)^{2}\right]^{-\beta_{\rm sw}},
\label{stellar}
\end{equation}
where $\beta_{\rm sw}=0.23$, $r_{\rm sw}=265$ pc, and $q_{\rm sw}=9.5\times10^{23}~\rm g~pc^{-3}yr^{-1}=3.2\times10^{-26}~\rm g~cm^{-3}Gyr^{-1}$. 
This particular profile was derived from the surface brightness of an elliptical galaxy \citep{lauer2007}.
We choose a larger $q_{\rm sw}$, compared to those used in \citet{perucho2014}, in order to consider the case where mass loading may lead to significant dynamical effects.

\subsection{Jet Setup}\label{s2.3}

{
As in Paper I, the computational domain is a 3D rectangular box for $z\!\ge\! 0$ that contains only the upward-moving jet; {no counterjet is explicitly included}. Jet material is injected along the $z$-axis through a circular nozzle located at the center of the bottom surface (see Figures \ref{fig1} and \ref{fig2} below).

The nozzle radius, $r_j$, determines the cross-sectional area of the jet inflow and sets the overall scale of the jet-induced structures.}
As mentioned above, in Table \ref{tab:t1}, two sets of models are considered: the ``r10'' models, with $r_j=10$~pc, present the early development of the jets on scales of $\sim 200-500$ pc, while the ``r100'' models, with $r_j=100$~pc, depict the later stage of the jets that extends up to $\sim 10$~kpc.

{The jets} are specified by the velocity, $v_j$, radius, $r_j$, density, $\rho_j$, and pressure, $p_j$, of the injected jet inflow. This set of the jet parameters are often translated into the jet power $Q_j$, the jet-to-background density contrast, $\eta\equiv\rho_j/\rho_c$,
and pressure contrast, $\zeta\equiv p_j/p_c$, for given background conditions ($\rho_c$ and $p_c$) \citep[e.g.,][Paper I]{rossi2020}. Here, the jet power is defined as
\begin{equation}
{Q_j} = \pi r_j^2 v_j \Big{(}\Gamma_j^2 \rho_j h_j - \Gamma_j \rho_j c^2\Big{)},\label{Qj}
\end{equation}
where $\Gamma_j=1/\sqrt{1-(v_j/c)^2}$ and $h_j$ are the {\it initial} bulk Lorentz factor and the specific enthalpy of the jet inflow, respectively. 
The jet parameters can also be combined to express the momentum injection rate or the jet thrust as \citep[e.g.,][Paper I]{perucho2014,hardcastle2020},
\begin{equation}
\dot{M}_j =\pi r_j^2 \left(\Gamma_j^{2}\rho_j\frac{h_j}{c^2}v_j^{2}+p_j\right).
\label{Mjet}
\end{equation}

\begin{deluxetable*}{cccccccccccccc}[t]
\tablecaption{Parameters of Jet Models$^a$\label{tab:t1}}
\tabletypesize{\small}
\tablecolumns{13}
\tablenum{1}
\tablewidth{0pt}
\tablehead{
\colhead{Model Name} &
\colhead{$Q_j$} &
\colhead{$r_j$} &
\colhead{$\eta$} &
\colhead{$\dot{M_j}$} &
\colhead{$\frac{v_j}{c}$} &
\colhead{$\Gamma_{j}$} &
\colhead{[$\langle\Gamma\rangle_{\rm{spine}}$] $^b$}&
\colhead{$\eta_r$} &
\colhead{$\frac{v_{\rm head}^*}{c}$} &
\colhead{[$\frac{v_{\rm head}}{c}$]$^c$} &
\colhead{$t_{\rm cross}$} &
\colhead{$\frac{t_{\rm end}}{t_{\rm cross}}$} \\
\colhead{}&
\colhead{($\rm{erg~s^{-1}}$)}&
\colhead{(pc)}&
\colhead{}&
\colhead{(dyne)}&
\colhead{}&
\colhead{}&
\colhead{}&
\colhead{}&
\colhead{}&
\colhead{}&
\colhead{(yr)}&
\colhead{}}
\startdata
	Q42-$\eta5$-r10 & 2.20E+42 &10 & 1.E-05  & 9.34E+31 & 0.95 & 3.2 & 2.07 & 1.27E-04  & 0.0106 & 0.005 & 3.07E+3  & 50  \\
	Q43-$\eta5$-r10 & 1.33E+43 &10 & 1.E-05  & 4.94E+32 & 0.99 & 7.1 & 4.42  & 6.28E-04  & 0.0242 & 0.014 & 1.35E+3  & 50  \\
	Q44-$\eta5$-r10 & 1.45E+44 &10 & 1.E-05 & 5.00E+33 & 0.999 & 22.4 & 8.68 & 6.25E-03  & 0.0732 &0.052  & 4.46E+2  & 50  \\
\hline
	Q42-$\eta5$-r10-P & 2.20E+42 &10 & 1.E-05 & 9.34E+31 & 0.95 & 3.2 & 2.08 & 1.27E-04  & 0.0106 &0.016 & 3.07E+3  & 28  \\
	Q42-$\eta5$-r10-M & 2.20E+42 &10 & 1.E-05 & 9.34E+31 & 0.95 & 3.2 & 1.68 & 1.27E-04  & 0.0106 &0.005 &  3.07E+3  & 50 \\
\hline
	Q44-$\eta5$-r100 & 3.50E+44 & 100 & 1.E-05 & 1.35E+34 & 0.9664 & 3.9 & 2.77 & 1.89E-04  & 0.0131 &0.024 & 2.49E+4 & 66  \\
	Q45-$\eta5$-r100 & 3.50E+45 &100 & 1.E-05  & 1.18E+35 & 0.996 & 11.2 & 5.65 & 1.55E-03  & 0.0379 & 0.089 & 8.61E+3 & 38 \\
	Q45-$\eta3$-r100 & 3.52E+45 &100 & 1.E-03  & 2.01E+35 & 0.85 & 1.9  & 1.82 & 3.61E-03 & 0.0482 & 0.146  & 6.77E+3  & 41 \\
\hline
\hline
\enddata
\tablenotetext{a}{Here, $\eta\equiv \rho_j/\rho_c$, $\Gamma_j\equiv [1-(v_j/c)^2]^{-1/2}$, $\eta_r \equiv (\rho_j h_j/\rho_c h_c)\Gamma_j^2$, and $v_{\rm head}^*\approx v_j\eta_r^{1/2}$. For all models, $\zeta\equiv p_j/p_c=1$.}
\tablenotetext{b}{The mean Lorentz factor estimated by averaging $\Gamma$ along the jet axis in the region of $z\leq2/3~l_{\rm{jet}}$, where the jet propagation length, $l_{\rm{jet}}$, is defined by the location of the contact discontinuity along the $z$-axis.}
\tablenotetext{c}{The jet-head advance speed estimated using the simulation data at $t_{\rm end}$.}
\end{deluxetable*}

For kinetically-dominated, light, relativistic jets with $v_j\approx c$, equations (\ref{Qj}) and (\ref{Mjet}) can be approximated as follows:
\begin{equation}
\frac{Q_j}{\pi r_j^2} \sim  \Gamma_j^2 \rho_j h_j v_j \sim \eta \Gamma_j^2 h_j \rho_c c \sim \eta_r \rho_c h_c c,
\label{Gammaj}
\end{equation}
\begin{equation}
\frac{\dot{M}_j}{\pi r_j^2} \sim \Gamma_j^{2}\rho_j h_j\left(\frac{v_j}{c}\right)^2 \sim \eta \Gamma_j^2 h_j \rho_c \sim  \eta_r \rho_c h_c.
\label{Mjet2}
\end{equation}
Here, the relativistic density contrast is given as
\begin{equation}
 \eta_r \equiv \frac{\rho_j h_j}{\rho_c h_c}\Gamma_j^2= \eta \Gamma_j^2  \left(\frac{h_j}{h_c}\right) \propto \eta \Gamma_j^2.  
\label{etar}
\end{equation} 
In this study, we assume $p_j=p_c$ (i.e., $\zeta=1$), hence $h_j= c^2 +(\epsilon_j + p_j)/\rho_j \sim c^2$.
Then, for identical background conditions, $\Gamma_j^2 \propto (Q_j/\eta \pi r_j^2)\propto (\dot{M}_j/\eta \pi r_j^2) \propto \eta_r/\eta $.
So, $\Gamma_j$ is larger for greater $Q_j/\pi r_j^2$ and for lighter jets with smaller $\eta$.

Initially, the jet propagation into the surrounding medium can be approximated by the advance speed of the jet head, $v_{\rm head}^* \approx v_j \cdot \sqrt{\eta_{r}} / (\sqrt{\eta_{r}} + 1)$ \citep{marti1997}. 
{This expression is derived from the balance between the jet's ram pressure and the background pressure in one-dimensional (1D) planar geometry. With the jet injected along the $z$-axis, this would provides a reasonable approximation for the {\it initial} velocity of the jet head.}
In our model setup with $\eta_r \ll 1$, the initial estimate for the jet advance speed can be approximated as
\begin{equation}
 \frac{v_{\rm head}^*}{c}\approx \left(\frac{v_j}{c}\right) \eta_r^{1/2} \approx \eta^{1/2}\Gamma_j \propto \left(\frac{Q_j}{\pi r_j^2}\right)^{1/2}.
\label{vhead}
\end{equation}
Again, for identical background conditions, the jet-head advance speed is faster for greater $(Q_j/\pi r_j^2)^{1/2}$.

Thus, the initial flow dynamics of our simulated jets are primarily determined by $\Gamma_j$ and $v_{\rm head}^*$, which depend on
the energy injection flux, $Q_j/\pi r_j^2$, and the density contrast, $\eta$, in our simulation setup with $\zeta=1$. 
In general, however, the characteristics of simulated jets are governed by four key parameters: $r_j$, defining the length scale; $Q_j$, specifying the energy input rate; $\eta$, specifying the density contrast; and $\dot{M}_j$, defining the momentum thrust of the jet inflow.

Table \ref{tab:t1} shows the ranges of these model parameters.
The first column lists the model name; following the nomenclature of Paper I, it includes three elements, the exponents of $Q_j$ and $\eta$, and $r_j$ in units of pc. We assign $\eta\equiv\rho_j/\rho_c=10^{-5}$ for the default setting. 
The three r10 models in the first group are the fiducial models, all sharing the same background profile described by Equation (\ref{kings}). 
In the second group, the Q42 model labeled ``P'' features a background profile following a $-3/2$ power-law, while the Q42 model labeled ``M'' incorporates mass loading from stellar winds. The r100 models in the third group are designed to explore the jet morphology on larger scales.

At the beginning, $\Gamma_j$ and $v_{\rm head}^*$, which are listed in the seventh and tenth columns of Table \ref{tab:t1}, control the flow dynamics. However, as the jet flow expands and decelerates, these two quantities gradually decrease with time.
For comparison, we present the mean Lorentz factor of the jet-spine flow, $\langle\Gamma\rangle_{\rm{spine}}$, and the actual jet-head advance speed, $v_{\rm head}$, in the eighth and eleventh columns, respectively. 
They are estimated using the simulation data at the end time of simulations, $t_{\rm end}$.
Additionally, in order to quantify the degree of the jet-flow deceleration, we define the ``deceleration factor'' 
as $\mathcal{R}_{\rm dec}\equiv v_{\rm head}/v_j$, which is listed in the {second} column of Table \ref{tab:t2}. 
It is worth noting that a smaller $\mathcal{R}_{\rm dec}$ indicates a stronger deceleration.
 
The jet crossing timescale is defined as $t_{\rm cross} \equiv {r_j}/{v_{\rm head}^*}$, and listed in the twelfth column of Table \ref{tab:t1}. The end time of simulations given in units of $t_{\rm cross}$ is listed in the thirteenth column. The spatial resolution is controlled by the number of grid zones across the initial jet radius, $N_j= r_j/\Delta x$. For all models, $N_j=8$, except for the Q44-$\eta5$-r100 model with $N_j=5$. Hence, for instance, {$(400)^3$ grid zones employed for the r10 models make up a box of volume $(500\ {\rm pc})^3$}.

The boundary condition of the jet nozzle plane ($z\!=\!0$) could affect the properties of simulated jets \citep[e.g.,][]{perucho2014,donohoe2016}. Here, we impose the continuous {\it outflow} condition to all the six faces of the simulation domain, including the $z\!=\!0$ plane, except at the jet nozzle.
The jet flow is injected mostly along the $z$-axis (i.e., $v_{z}\approx v_j$); yet, we introduce a slow, small-angle precession with period $\tau_{\rm prec}=3~t_{\rm cross}$, and angle $\theta_{\rm prec}=1^{\circ}$, to break the rotational symmetry (see Paper I for details).

\subsection{Modeling for Synchrotron Surface Brightness}\label{s2.4}

{To generate synthetic maps of simulated jets, we estimate the synchrotron emission adopting simple modelings:} 
we utilize physically motivated prescriptions for the magnetic field strength, $B^{\prime}$, and the cosmic-ray electron population, $\mathcal{N}^{\prime}_{e}(\gamma_e^{\prime})$, which are not explicitly modeled within our RHD simulations.
Here, $\gamma_e^{\prime}$ is the Lorentz factor of cosmic ray (CR) electrons. Throughout the paper, the primed variables, such as $B^{\prime}$ and $\gamma_e^{\prime}$, represent the quantities defined in the local fluid frame, while unprimed variables are used for the quantities defined in the computational or observer frame.

\subsubsection{Magnetic Field Models}\label{s2.4.1}

The details of magnetic field modeling were provided in \citet{seo2023,seo2024}; therefore, here we offer only a brief overview of the underlying physics. We consider two well-known MHD processes that can amplify magnetic fields: small-scale turbulent dynamo with saturated magnetic energy density $\mathcal{E}_{B,{\rm turb}}\equiv B_{\rm{turb}}^2/8\pi\approx\mathcal{E}_{\rm{turb}}$ \citep[e.g.,][]{cho2009}
and CR streaming instabilities resulting in $\mathcal{E}_{B,{\rm Bell}}\equiv B_{\rm{Bell}}^{2}/8\pi\approx(3/2)(u_s/c)P_{\rm{CRp}}$ at quasi-parallel shocks \citep[e.g.,][]{bell2004}. 
Here, $\mathcal{E}_{\rm turb}\approx \Gamma_{\rm turb}\left(\Gamma_{\rm turb}-1\right)\rho c^2$ is estimated using the turbulent flow speed, $u_{\rm turb}$. 
The pressure of CR protons, accelerated via diffusive shock acceleration at shocks with the speed, $u_s$, and the preshock density, $\rho_1$, is set to be $P_{\rm{CRp}}\sim0.1\rho_1u_s^2$ \citep{caprioli2014b}.
In the regions of shock-free and weak turbulence, $P_{B,P}\equiv B_P^2/8\pi\approx p/\beta_p$ with $\beta_p=100$ is adopted.
We estimate $u_{\rm{turb}}$, $u_s$, $\rho_1$, and $p$ using the hydrodynamic variables from the simulated jet flows. 
The highest estimate among the three model values, $B^{\prime}={\rm{max}}(B_{\rm{turb}},B_{\rm{Bell}},B_P)$, is selected as the local comoving magnetic field strength.

{Applying these prescriptions to the RHD simulation data results in a magnetic field distribution determined by $B_{\rm{turb}}$ in the turbulent regions of the jet spine and the cocoon and by $B_P$ in the relatively quiescent regions such as the shocked background medium.
The plasma beta in the ISM and ICM, however, has a ranges of values, for instance, $\beta_p\approx1-100$.} We here adopt $\beta_p=100$ uniformly across all our models, as our primary focus is on the morphology of radio images of the jet spine and the cocoon.
In fact, we find that the radio images of our simulated jets exhibits only weak dependence on the adopted value of $\beta_p$. 

\subsubsection{Cosmic-Ray Electron Population}\label{s2.4.2}

The jet ejected from the central engine is thought to be composed of ionized plasma, which includes relativistic electrons, protons, and possibly positrons \citep[e.g.,][]{begelman1984}.
Electrons are expected to undergo further acceleration to CRs, due to shocks and MHD turbulence within the jet-induced flows \citep[e.g.,][]{bell1978,drury1983, brunetti2007}. While a comprehensive theoretical modeling of all the relevant processes would be quite complex, 
we adopt a simplified phenomenological approach available in the literature, for instance, \citet{gomez1995} to model the CR electron population. The number and energy densities of CR electrons in the fluid frame are assumed to be proportional to the rest-mass density $\rho$ and the internal energy density $\epsilon$ of the fluid, respectively. In addition, the energy spectrum of CR electrons is represented by a power-law form, as follows:
\begin{equation}
\mathcal{N}^{\prime}_{e}(\gamma_e^{\prime}) = \mathcal{N}_{0}^{\prime}(x,y,z)\cdot \gamma_e^{\prime-\sigma},\label{NeCR}
\end{equation}
where the energy of CR electrons is given by the Lorentz factor, $\gamma_e^{\prime}$.  
We employ a single power-law slope of $\sigma=2.2$, which corresponds to the representative slope for the energy spectrum of CR electrons accelerated at relativistic shocks. Electron cooling due to synchrotron and inverse-Compton losses is not considered.

The normalization factor, $\mathcal{N}^{\prime}_{0}$, depends on $\rho$ and $\epsilon$ of the local fluid as follows:
\begin{equation}
\mathcal{N}_{0}^{\prime} \propto f_{\rm jet}\cdot \Big{(} \frac{\epsilon(\sigma-2)}{1-C_{E}^{2-\sigma}} \Big{)}^{\sigma-1} \Big{(} \frac{1-C_{E}^{1-\sigma}}{\rho(\sigma-1)} \Big{)}^{\sigma-2},\label{Ne0}
\end{equation}
where $C_{E}=\gamma_{\rm max}/\gamma_{\rm min}$ is the ratio of the maximum and minimum energies of CR electrons \citep{gomez1995}.
This parameter is kept constant at $C_{E}=10^{3}$ (e.g., $\gamma_{\rm min}\approx 10^3$ and $\gamma_{\rm max} \approx 10^6$) for all models.
An ad-hoc weight function, $f_{\rm jet}=9(5/3-\gamma)^2$, is introduced in equation (\ref{Ne0}) to assign the CR electron population preferentially to the relativistic plasma contained in the jet flow: for relativistic fluid with $\gamma=4/3$, $f_{\rm jet}=1$, while for thermal fluid with $\gamma=5/3$, $f_{\rm jet}=0$.
{We point out that inside the cocoon, the relativistic electrons of the jet material are mixed with the nonrelativistic thermal electrons of the background medium, causing the adiabatic index of the fluid to range between $1.47\lesssim\gamma\lesssim5/3$ in our jet models (see Figure \ref{fig3}).}
In our model calculations, the normalization factor scales as ${N}_{0}^{\prime}\propto f_{\rm jet}\cdot \epsilon(\epsilon/\rho)^{\sigma-2}$, while its absolute amplitude is arbitrary.

\begin{figure*}[t]
\centering
\includegraphics[width=1\linewidth]{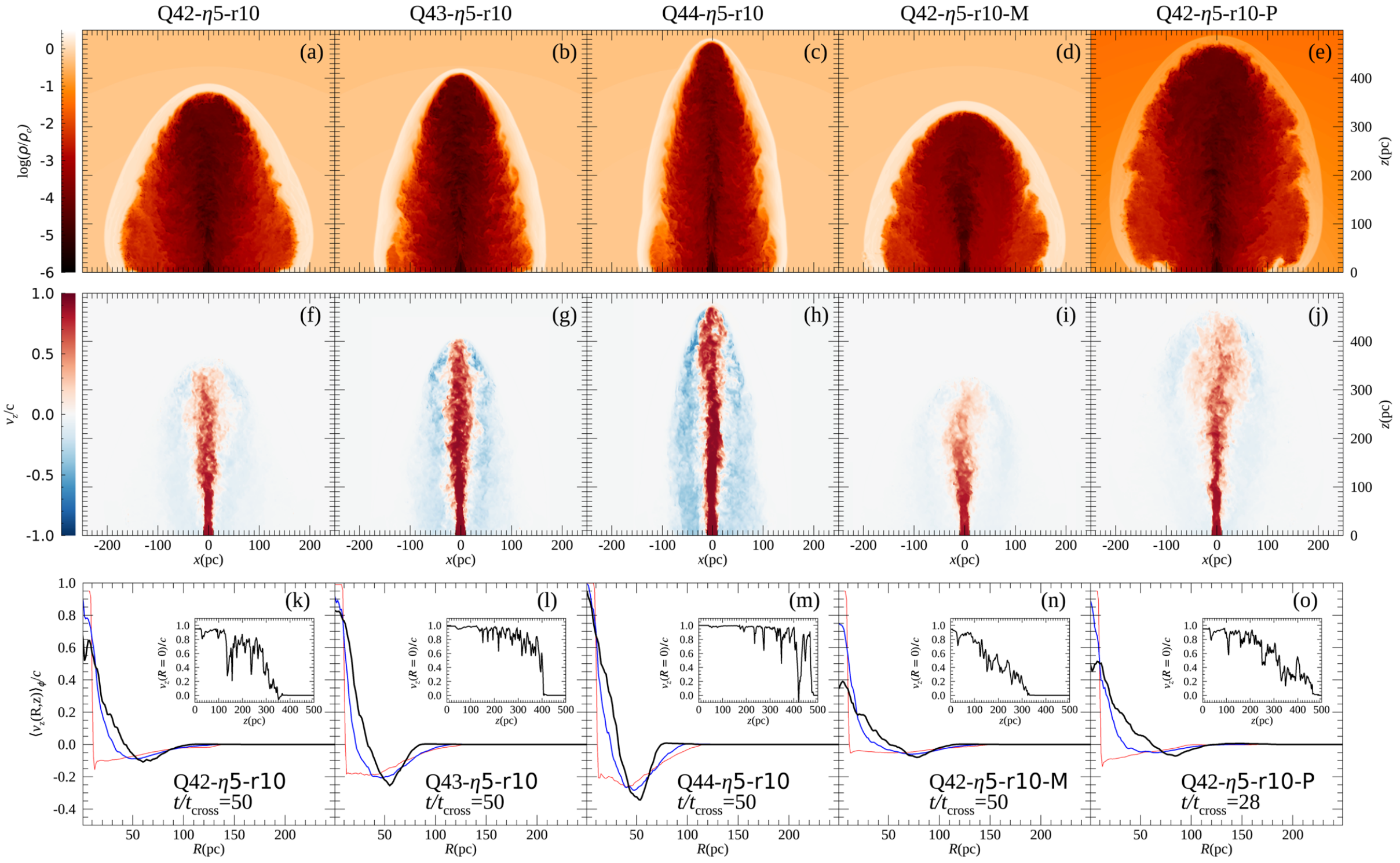}
\caption{Structures of the jet-induced flows in the r10 models, Q42-$\eta5$-r10, Q43-$\eta5$-r10, Q44-$\eta5$-r10, Q42-$\eta5$-r10-M, and Q42-$\eta5$-r10-P, at the end of simulations. See Table \ref{tab:t1} for the model parameters. (a-e) 2D slice images of $\log(\rho/\rho_{c})$ in the $x$-$z$ plane through $y\!=\!0$.  (f-j) 2D slice images of the vertical velocity, $v_{z}/c$, in the same plane. (k-o) The vertical velocity averaged over the azimuthal angle $\phi$, $\langle v_z(R,z)/c \rangle_{\phi}$, at three different $z$'s: $z\!=\!0$ (red), $z_1\equiv l_{\rm jet}/3$ (blue), and $z_2\equiv2l_{\rm jet}/3$ (black). Here $R\!=\!(x^2+y^2)^{1/2}$ is the transverse distance from the $z$-axis, and $l_{\rm jet}$ is the jet propagation length. { The inset of each panel of (k-o) shows the vertical velocity along the jet axis, $v_z(R\!=\!0)/c$, across the $z$ distance.}}
\label{fig1}
\end{figure*}

\subsubsection{Synchrotron Emission}\label{s2.4.3}

The synchrotron power emitted by the CR electrons in equation (\ref{NeCR}) with an isotropic pitch angle distribution can be estimated as follows \citep{rybicki1979}:
\begin{equation}
P^{\prime}_{\nu^{\prime}}= \frac{2 \pi \sqrt{3} q_e^2\nu_L} {c} \int_{\gamma_{\rm min}}^{\gamma_{\rm max}} d\gamma_e^{\prime} \mathcal{N}_{0}^{\prime}\gamma_e^{\prime-\sigma}\cdot  \xi \int_{\xi}^{\infty} K_{5/3}(\chi)d\chi, 
\label{Pnu}
\end{equation}
where $\nu_L= q_e B^{\prime} /2\pi m_e c$ is the Larmor frequency, $\nu_c =(3/2)\gamma_e^{\prime 2} \nu_L$ is the critical frequency, $q_e$ is the electron charge, $m_e$ is the electron mass, $\xi\equiv \nu^{\prime}/\nu_c$, and $K_{5/3}(\xi)$ is a modified Bessel function.
For $\gamma_{\rm max}\gg \gamma_{\rm min}\gg 1$, the synchrotron volume emissivity can be expressed as
\begin{equation}
j^{\prime}_{\nu^{\prime}} = \frac{P^{\prime}_{\nu^{\prime}}}{4\pi} \propto f_{\rm jet}\cdot \epsilon \left(\frac{\epsilon}{\rho}\right)^{\sigma-2} \cdot B^{\prime(\sigma+1)/2} \nu^{\prime-(\sigma-1)/2}.
\end{equation}
{Then, the observed intensity, or surface brightness, can be calculated by integrating along the LOS. 
For example, when the $y$-axis represents the LOS, the intensity is given by $I_{\nu}(x,z) = \int j_{\nu}(x,y,z)dy$.}

In the computational frame, the observed frequency shifts to $\nu=\mathcal{D}{\nu^{'}}$, where $\mathcal{D}=1/(\Gamma[1-(v/c) \cos\theta])$ is the Doppler factor, and $\theta$ is the angle between the fluid velocity vector, $\Vec{v}$, and the line-of-sight (LOS). With relativistic beaming, the volume emissivity is given as $j_{\nu} = \mathcal{D}^{2}j^{'}_{\nu^{'}}(\nu/\mathcal{D})$ in the computational frame.
For ``observed'' images,
the inclination angle, $\theta_{\rm obs}$, with respect to the jet axis (see Figure \ref{fig7}(e)), {if nonzero,} could have important consequences.
It controls {not only} the relativistic Doppler factor $\mathcal{D}$, {but also} the number of bright points with high $j^{'}_{\nu^{'}}$ values along the path of integration. To incorporate such effects, we tilt the simulation box by the angle of $\pi/2-\theta_{\rm obs}$ around the $x$-axis, {and then calculate the integration along the LOS.}

We ignore the synchrotron self-absorption because we expect it to be insignificant in the diffuse plasmas in radio jets. 

\begin{figure}[t]
\centering
\includegraphics[width=1.0\linewidth]{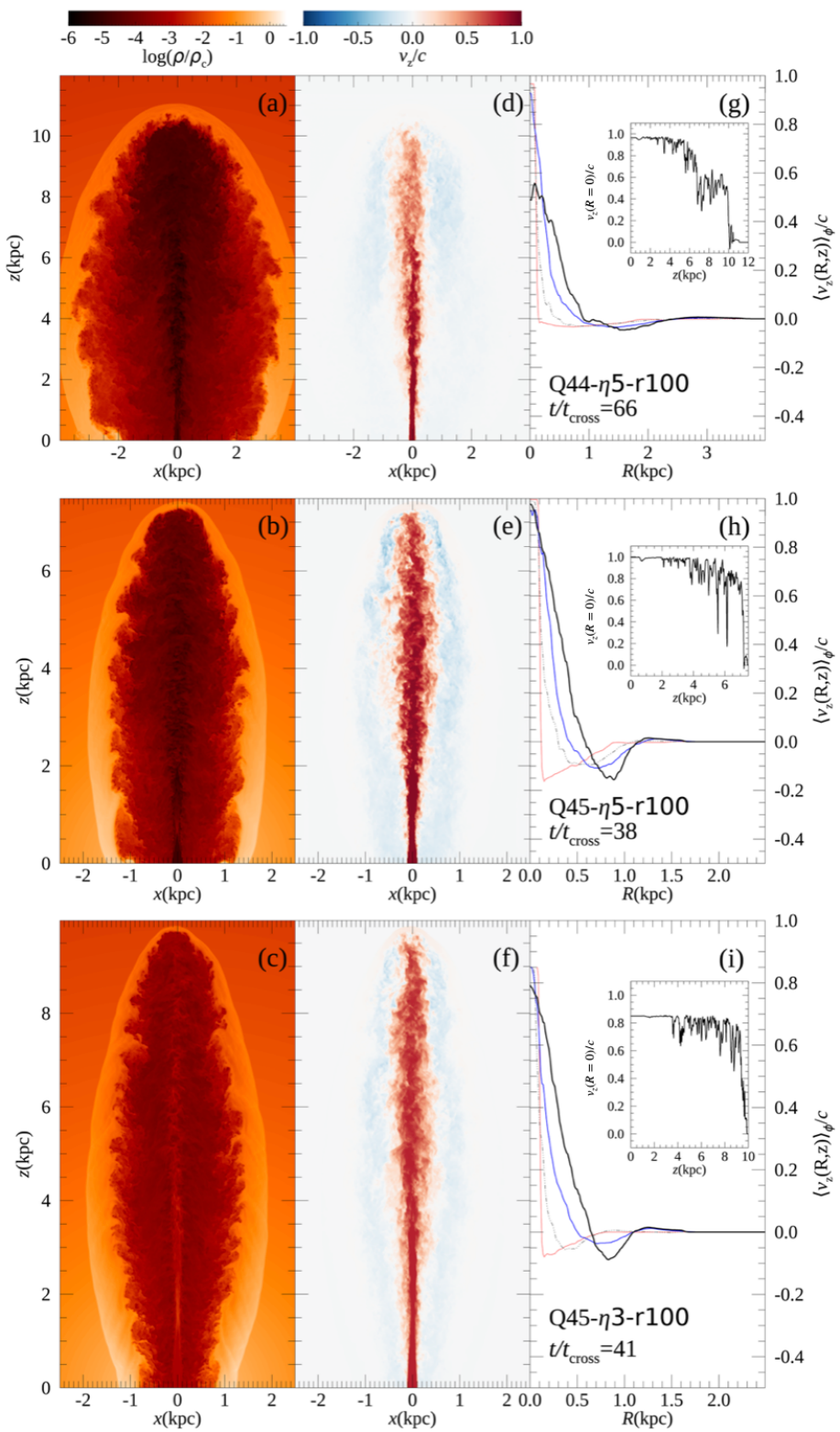}
\caption{Same as Figure \ref{fig1}, except for the r100 models, Q44-$\eta5$-r100, Q45-$\eta5$-r100, and Q45-$\eta3$-r100, shown. See Table \ref{tab:t1} for the model parameters.}
    \label{fig2}
\end{figure}

\section{Structures and Dynamics of Jet-Induced Flows}\label{s3}

\subsection{Flow Structures of Jets}\label{s3.1}

Figures \ref{fig1} and \ref{fig2} display 2D slice images of the density, $\log(\rho/\rho_c)$, and the vertical components of the flow velocity, $v_{z}/c$, along with 1D profiles of the vertical velocity averaged over the azimuthal angle $\phi$, $\langle v_z(R,z)/c \rangle_{\phi}$, across the transverse $R\!=\!(x^2 + y^2)^{1/2}$ direction for the five r10 models and three r100 models, respectively (see Table \ref{tab:t1}). {In addition, the vertical velocity along the jet axis, $v_z(R\!=\!0,z)/c$, is displayed across the $z$ distance in the small insets.}

{Consistent with earlier numerical studies of FR-I jets mentioned in the introduction}, the jets in Figures \ref{fig1} and \ref{fig2} have the following parts in common: the bow shock, the shocked background medium, the backflow, and the jet-spine flow. 
The backflow is identifiable as blue regions in the 2D distribution of $v_{z}(x,z)/c$, while the jet-spine flow appears red. Surrounded by the bow shock, the shocked background medium would appear as the surface of the X-ray cavity in observation. Inside, the plume-like cocoon, filled with turbulence, develops. The shape of the cocoon changes from laterally expanded to longitudinally elongated, as the jet power increases. Significant ``mixing'' is observed between the jet-spine flow and the backflow, as well as between the cocoon material and the shocked background medium. Instabilities, such as Kelvin-Helmholtz instability at the shear interfaces, contribute to the mixing. In the lower-power models, the mixing at the boundary between the jet-spine flow and the backflow is more extended, inducing wider cocoons filled with more substantial turbulence. We note that the evolution timescale, $t_{\rm cross}$, is shorter for higher $v_{\rm head}^{*}$, as listed in Table \ref{tab:t1}.

\begin{figure*}[t]
\centering
\includegraphics[width=0.9\linewidth]{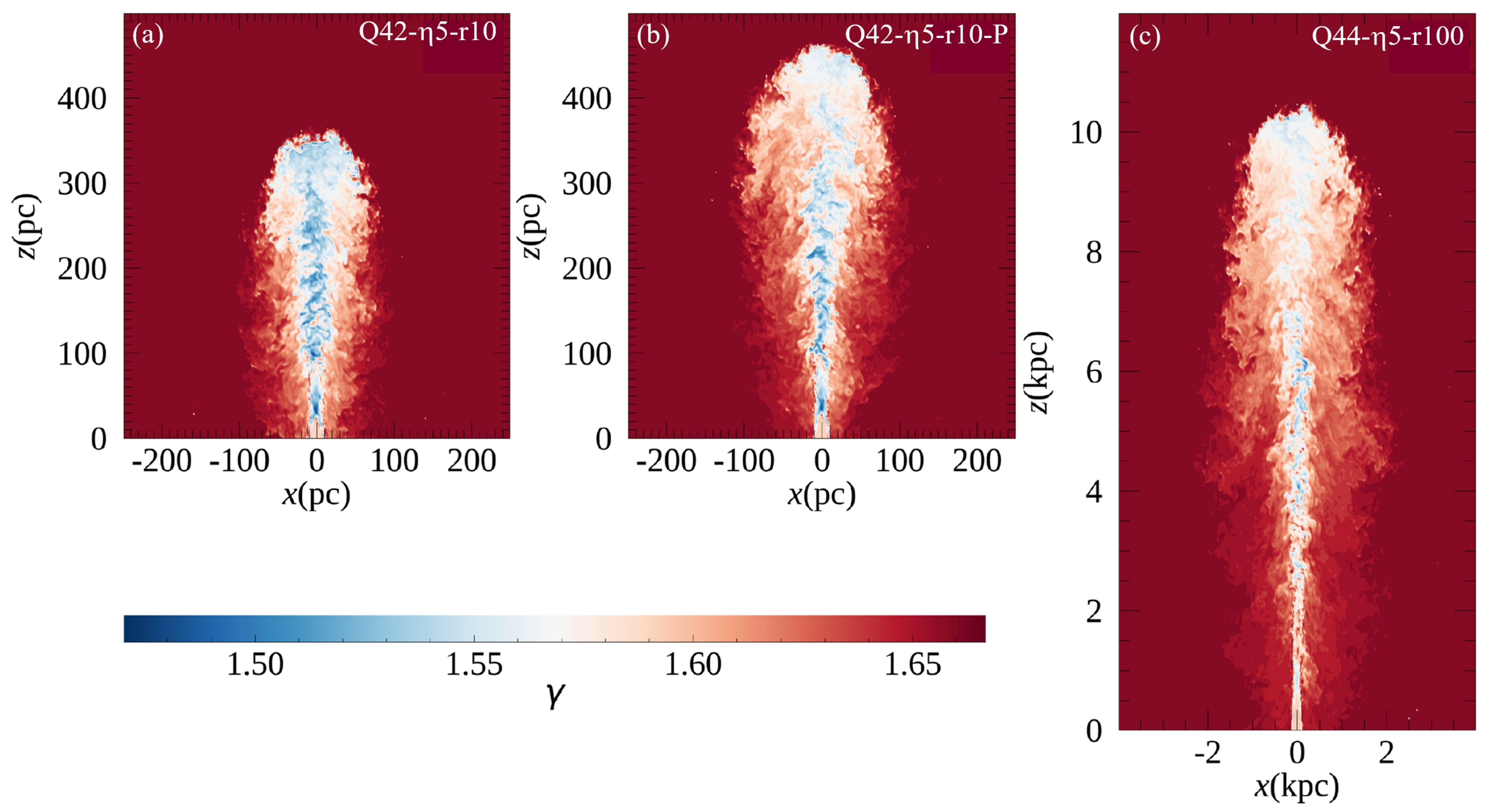}
\caption{{2D slice images of the adiabatic index, $\gamma$, for Q42-$\eta5$-r10, Q42-$\eta5$-r10-P, and  Q44-$\eta5$-r100.  In these models, the jet inflow is hotter than the background, with $T_j \approx T_b/\eta$ and $\gamma_j \approx 1.6$ at injection. After passing through shocks, the jet fluid becomes thermally relativistic, with $\gamma \approx 1.47$ (dark blue), while the background gas remains non-relativistic with $\gamma = 5/3$ (dark red). In the backflow region, shown primarily in a whitish peach hue, the jet material mixes with the ambient gas, mainly through the Kelvin-Helmholtz instability.}}
    \label{fig3}
\end{figure*}

{
The jet flow is kinematically relativistic at injection, but its specific enthalpy is only {mildly} relativistic with $h_j/c^2\approx1.3$, and its adiabatic index is $\gamma_j\approx 1.6$, in our jet models. However, the kinetic energy is converted to the internal energy in the post-shock regions of relativistic shocks, including recollimation shocks.
As a result, the fluid's adiabatic index decreases, as shown in Figure \ref{fig3}. The formation and growth of turbulent shear layers mix the shocked jet material (dark blue) with the ambient gas that has nonrelativistic enthalpy (dark red). We see the mixing to be more efficient beyond recollimation shocks at 200 pc for the Q42-$\eta5$-r10-P case and beyond 6 kpc for the Q44-$\eta5$-r100 case.}

The r10 models {produce} the jet-induced flows confined within core regions ($r\!<\!r_c\!=\!1.2$~kpc) of typical elliptical galaxies, possibly reflecting the flow structures of typical radio jets in their early phase \citep{perucho2007,perucho2014}. In Figures \ref{fig1}(a-c) and (f-h), the three fiducial models, Q42-$\eta5$-r10, Q43-$\eta5$-r10, and Q44-$\eta5$-r10, demonstrate how the jet structures evolve depending on the jet power; higher $Q_j$ (or higher $v_{\rm head}^{*}$) leads to faster penetration of the jet head into the background medium, resulting in a more elongated cocoon. These findings align closely with previous studies \citep[e.g.,][]{marti1997,rossi2008,perucho2007,li2018}.

\begin{figure}[t]
\centering
\vskip 0.5 cm
\includegraphics[width=1\linewidth]{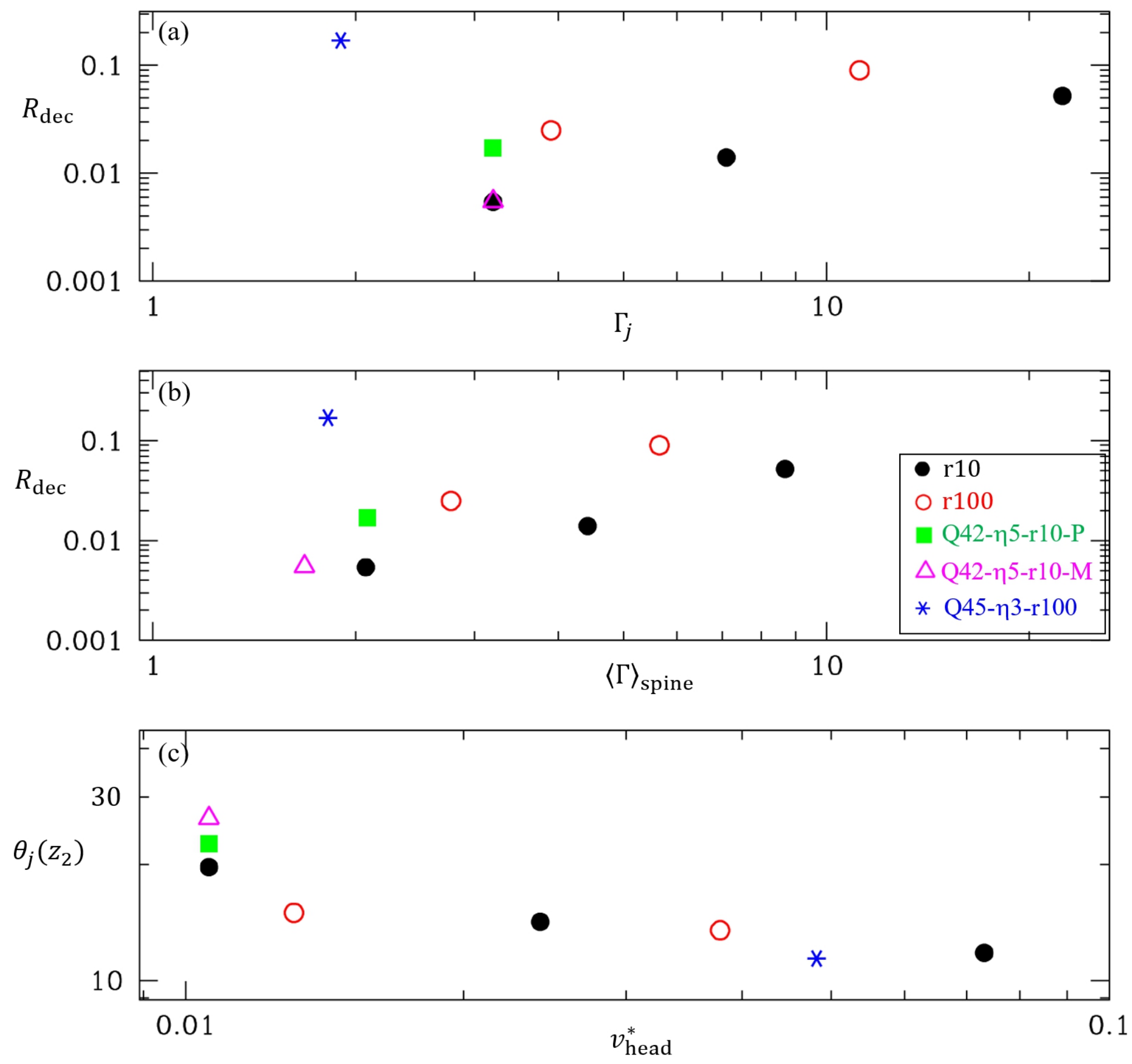}
\caption{
(a) Deceleration factor, $\mathcal{R}_{\rm dec}\equiv v_{\rm head}/v_j$, versus the initial Lorentz factor of the jet, $\Gamma_j$.
(b) $\mathcal{R}_{\rm dec}$ versus the mean Lorentz factor of the jet-spine flow, $\langle\Gamma\rangle_{\rm{spine}}$.
(c) Jet-spine spreading angle at $z_2$, $\theta_j(z_2)$, versus the initial jet-head speed, $v_{\rm head}^*$. 
All the eight jet models listed in Tables \ref{tab:t1} are shown with different symbols. }
\label{fig4}
\end{figure}

The red regions with $v_z>0$ in Figures \ref{fig1}(f-j) show the degree of the jet-head spreading, which is governed by the relative flow speeds in the longitudinal and transverse directions. We define the jet-head spreading angle as $\theta_j(z)\equiv 2 \arctan [\Delta R(z)/z]$, where $\Delta R(z)$ is the transverse width of the jet-spine region with $v_z>0$ and $z$ is the distance from the injection point. 
The second column of Table \ref{tab:t2} lists the value of $\theta_j(z_2)$ at $z_2\!=\! 2l_{\rm jet}/3$,
where the length of the jet, $l_{\rm jet}$, is the distance between the injection point and the contact discontinuity along the $z$-axis in the simulated jet models at $t_{\rm end}$.
Among the five models shown in Figure \ref{fig1}, the jet-head spreading is larger in models with lower $Q_j$ because the longitudinal deceleration is greater for lower $\Gamma_j$. 

Figure \ref{fig4}(c) shows the overall {relation} between $\theta_j(z_2)$ and $v_{\rm head}^*$. 
It is the largest in Q42-$\eta5$-r10-M (magenta triangle), in which the mean Lorentz factor of the jet-spine flow, $\langle\Gamma\rangle_{\rm{spine}}$, is the smallest due to the mass loading. 
On the other hand, $\theta_j(z_2)$ is slightly larger in Q42-$\eta5$-r10-P (green square) than that in Q42-$\eta5$-r10, due to the {larger} transverse spreading of the jet in the {power-law gradient of the external density/pressure}. 
Section \ref{s3.2} discusses the last two cases in detail.

\begin{deluxetable*}{lcccccccc}[ht]
	\tablecaption{Morphological Parameters for Simulated Jets\label{tab:t2}}
\tabletypesize{\small}
\tablecolumns{10}
\tablenum{2}
\tablewidth{0pt}
\tablehead{
\colhead{Model name} &
        \colhead{[$\mathcal{R}_{\rm dec}$]$^a$}&
        \colhead{[$\theta_j(z_2)$] $^b$} &
        \colhead{{[$v_z(R\!=\!0,z_2)/c$] $^c$}} &
        \colhead{[$\log \mathcal{F}|_{\theta_{\rm obs}=90^{\circ}}$] $^d$} &
	\colhead{[$\frac{d_b}{d_{f}}|_{\theta_{\rm obs}=90^{\circ}}$] $^e$} &
	\colhead{$\frac{d_b}{d_{f}}|_{\theta_{\rm obs}=60^{\circ}}$} &
	\colhead{$\frac{d_b}{d_{f}}|_{\theta_{\rm obs}=30^{\circ}}$} &
}
\startdata
	Q42-$\eta5$-r10 & 5.4E-3  & 19.7 & 0.75 & 0.0 & 0.94 & 0.81(0.88) & 0.24(0.79)\\
	Q43-$\eta5$-r10 & 1.4E-2 & 14.2 & 0.89 & 1.7 & 0.98 & 0.95 & 0.93\\
	Q44-$\eta5$-r10 & 5.2E-2 & 11.8 & 0.96 & 4.1 & $\sim1$ & $\sim1$ & 0.93\\
\hline
        Q42-$\eta5$-r10-P & 1.7E-2 & 22.7 & 0.46 & -0.4 & 0.47 & 0.45(0.92) & 0.07(0.62)\\
        Q42-$\eta5$-r10-M & 5.5E-3 & 26.4 & 0.35 & -0.6 & 0.25 & 0.09 & 0.07\\
\hline
	Q44-$\eta5$-r100 & 2.5E-2 & 15.0 & 0.40 & -0.1 & 0.58 & 0.61(0.91) & 0.34(0.77)\\
	Q45-$\eta5$-r100 & 9.0E-2 & 13.5 & 0.98 & 2.9 & 0.98 & 0.96 & 0.89\\
	Q45-$\eta3$-r100 & 1.7E-1 & 11.4 & 0.76 & 2.4 & 0.96 & 0.98 & 0.96\\
\hline
\hline
\enddata
\tablenotetext{a}{The deceleration factor is defined as $\mathcal{R}_{\rm dec}\equiv v_{\rm head}/v_j$; a smaller $\mathcal{R}_{\rm dec}$ indicates a stronger deceleration.}
\tablenotetext{b}{The jet-spine spreading angle, $\theta_j(z_2)\equiv 2 \arctan [\Delta R(z_2)/z_2]$, at $z=z_2\equiv2l_{\rm jet}/3$. Here, $\Delta R(z_2)$ is the transverse width of the jet-spine region with the positive vertical velocity, $v_z>0$, at $z=z_2$.}
{\tablenotetext{c}{$v_z(R\!=\!0,z_2)/c$ is the vertical velocity along the jet axis at $z=z_2$.}}
\tablenotetext{d}{$\mathcal{F}|_{\theta_{\rm obs}=90^{\circ}}= I_{\nu}^{\rm max}/I_{\nu,N}^{\rm Q42-\eta5-r10}$, where $I_{\nu}^{\rm max}$ is the peak value in the face-on map ($\theta_{\rm obs}=90^{\circ}$) for each model and $I_{\nu,N}^{\rm Q42-\eta5-r10}$ is the peak value in the face-on map of the Q42-$\eta5$-r10 model.}
\tablenotetext{e}{The ratio of the projected distances to the brightest spot, $d_{b}$, and the faintest edge, $d_{f}$, obtained from the surface brightness maps of Figures \ref{fig8}-\ref{fig9}. The values for the counterjets are also given in the parenthesis for selected models.}
\end{deluxetable*}

In Figures \ref{fig1}(k-m), the profile of $\langle v_z(R,z)/c \rangle_{\phi}$ across the transverse $R$ direction displays a stronger deceleration in the models with lower $Q_j$, whereas deeper dips with negative values indicate faster downward-moving velocity of the backflow in the models with higher $Q_j$. The velocity curves at different $z$ locations also show spreading of upward-moving velocity with increasing $z$, implying stronger decollimation in the models with lower $Q_j$. The fourth column of Table \ref{tab:t2} also confirms that the vertical velocity at $z_2\!=\!2l_{\rm jet}/3$ in the $z$-axis, {$v_z(R\!=\!0,z_2)/c$}, increases with increasing $Q_j$. Such trends of deceleration and decollimation agree with the visual impression of the 2D distribution of $v_z(x,z)$ displayed in Figures \ref{fig1}(f-h).

In the insets of Figures \ref{fig1}(k-m), the vertical velocity along the jet axis, {$v_z(R\!=\!0)/c$}, exhibits a substantial deceleration in the models with low $Q_j$. This deceleration results from mixing at the jet-backflow interface and the entrainment of the background medium.  
Among the {fiducial} r10 models, the deceleration factor is the smallest, at $\mathcal{R}_{\rm dec}=5.3\times 10^{-3}$ in the Q42 case, indicating the most significant deceleration. In contrast, $\mathcal{R}_{\rm dec}$ is largest, at $5.0\times 10^{-2}$ in the Q44 case.
Figures \ref{fig4}(a-b) illustrate that, overall, $\mathcal{R}_{\rm dec}$ is smaller (i.e., stronger deceleration) for smaller $\Gamma_j$ or smaller $\langle\Gamma\rangle_{\rm{spine}}$ (see also Table \ref{tab:t1}).

As a result, in the low-power Q42 models, the vertical flow diffuses relatively smoothly toward the jet head (see Figure \ref{fig1}(f)), while in the high-power Q44 model, it comes to an abrupt stop {at the jet head} (see Figure \ref{fig1}(h)). The mean Lorentz factor of the jet-spine flow is estimated as $\langle\Gamma\rangle_{\rm{spine}}\approx 2.1$, 4.4, and 8.7 for the Q42, Q43, and Q45 models, respectively (see Table \ref{tab:t1}). Overall, the vertical velocity of the backflow is faster for higher $Q_j$: $|v_z|\sim0.1c$ for the Q42 models and $|v_z|\sim 0.5c$ for the Q44 model. 

Along the jet-spine flow, multiple recollimation shocks appear. In each model, the first recollimation shock is located at the initial dip in the {$v_z(R\!=\!0)/c$} profile, which is deepest in the Q42-$\eta5$-r10-M model and shallowest in the Q44-$\eta5$-r10 model. This reflects that the induced shocks have higher Mach numbers {(or higher speeds in the shock rest frame)} in higher power jets. A few additional recollimation shocks follow along the jet-spine flow. These shocks are also manifested as discontinuous jumps in the 2D distribution of $\rho$ and $v_z$. Apart from the oscillations arising from instabilities and jet precession, the recollimation shocks are almost stationary in the computational frame for a steady jet inflow.

\begin{figure*}[t]
\centering
\includegraphics[width=0.9\linewidth]{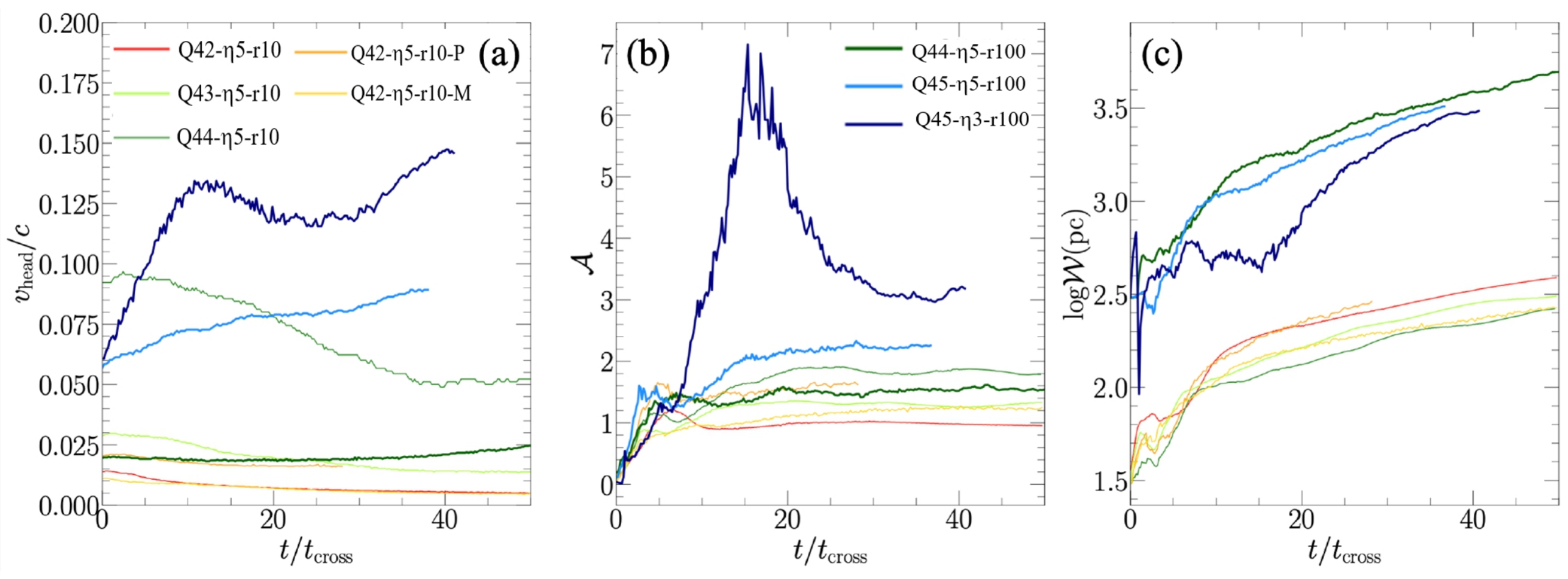}
\caption{
Time variations of (a) the jet-head advance speed, $v_{\rm head}$, (b) the axial ratio, $\mathcal{A}=l_{\rm jet}/\mathcal{W}$, and (c) the lateral width, $\mathcal{W}$, of the lobe for all the models listed in Table \ref{tab:t1}.
The r100 models that generate the jet flows propagating beyond the galactic core ($\sim \rm kpc$) are denoted with bold lines.}
    \label{fig5}
\end{figure*}

\subsection{Effects of Mass Loading and Environment}\label{s3.2}

\subsubsection{Q42-r10 Models in Different Environment}

In Figure \ref{fig1}, the comparison of Q42-$\eta5$-r10 and Q42-$\eta5$-r10-M (the model with mass loading) shows how mass loading affects the development of the jet-induced flows. 
The density distribution in Figure \ref{fig1}(d) shows that in the Q42-$\eta5$-r10-M model, the mixing between the jet spine and shear layers starts earlier. There is also a stronger recollimation shock at $\sim30 \rm pc$, a shorter cocoon with the jet length $l_{\rm jet}\approx 330~\rm pc$, and more complex, unstable structures along the shear interfaces, compared to Q42-$\eta5$-r10.
The vertical velocity profile in Figures \ref{fig1}(i) and (n) show a more substantial deceleration of the jet propagation due to mass loading, resulting in a slower downward-moving backflow. While $\langle\Gamma\rangle_{\rm{spine}}\approx 2.1$ for Q42-$\eta5$-r10, it decreases to 1.7 for Q42-$\eta5$-r10-M.

The difference between Q42-$\eta5$-r10 and Q42-$\eta5$-r10-P shows how the background density with a power-law decline, $\rho_b(r)\propto r^{-3/2}$, affects the flow dynamics. In Q42-$\eta5$-r10-P, the cocoon grows more rapidly after the jet head propagates beyond $\sim200~\rm pc$, where the flow becomes transverse-expansion-dominated. As shown in panels (e), (j), and (o) of Figure \ref{fig1}, owing to the decreasing pressure away from the center, the jet spine propagates faster, and the cocoon expands faster in both the longitudinal and transverse directions, compared to the fiducial model Q42-$\eta5$-r10. The length of the jet extends to approximately $l_{\rm jet} \sim470~\rm pc$ at $28~t_{\rm cross}$ in Q42-$\eta5$-r10-P, while $l_{\rm jet} \sim370~\rm pc$ at $50~t_{\rm cross}$ in Q42-$\eta5$-r10. In Q42-$\eta5$-r10-P, the deceleration factor, $\mathcal{R}_{\rm dec}=1.6\times 10^{-2}$, is three times larger than that of the fiducial model, meaning less deceleration.  Despite such differences, the mean Lorentz of the jet spine, $\langle\Gamma\rangle_{\rm{spine}}\approx 2.1$, is similar to that of the fiducial model.

\subsubsection{Q44 Models on Different Scale}

{Q44-$\eta5$-r100 with $r_j=100$~pc (Figure \ref{fig2}) and Q44-$\eta5$-r10 with $r_j=10$~pc (Figure \ref{fig1}) have the same power but differ in radius. The models vary in two key aspects:} (1) In the r100 model, both $Q_j/\pi r_j^2$ and $\dot{M}j/\pi r_j^2$ are lower than in the r10 model, which leads to smaller $\Gamma_j$ and $v_{\rm head}^*$ (see Table \ref{tab:t1}). (2) In the r100 model, the jet propagates through the background medium, encountering decreasing pressure once it surpasses the core radius ($r>r_c$).
These two factors influence the deceleration of the jet flow in opposite directions: smaller $\Gamma_j$ and smaller $v_{\rm head}^*$ lead to a stronger deceleration, whereas declining background pressure results in faster expansion.

As a result, in the Q44-$\eta5$-r100 model, $\langle\Gamma\rangle_{\rm{spine}}\approx 2.8$ and $\mathcal{R}_{\rm dec}\approx 2.6\times 10^{-2}$; whereas  in the Q44-$\eta5$-r10 model, $\langle\Gamma\rangle_{\rm{spine}}\approx 8.7$ and $\mathcal{R}_{\rm dec}\approx 5.0\times 10^{-2}$, with the jet contained within a nearly uniform core. Due to the loss of pressure confinement after crossing the core, the jet expands more rapidly in the transverse direction, causing the jet flow to slow down more significantly in Q44-$\eta5$-r100 compared to Q44-$\eta5$-r10.

The ambient density/pressure given in equation (\ref{kings}) asymptotically approaches a power-law distribution, $\sim r^{-\beta_a}$, where $\beta_a = 3\beta_K \sim 2.19$ somewhere beyond $r > r_c$. It was argued through analytical calculations that for $\beta_a>2$, the jet can become {more or less} free to expand due to the loss of the pressure confinement \citep{bromberg2011}. We confirm that through the bulging and expansion of the r100 jets.

\subsubsection{Q45-r100 Models with Different Density Contrasts}

The difference between Q45-$\eta5$-r100 and Q45-$\eta3$-r100, as shown in the second and third rows of Figure \ref{fig2}, respectively, highlights the dependence of jet flow dynamics on the jet-to-background density contrast.
{In the Q45-$\eta3$-r100 model, where $\rho_j$ is 100 times higher, the energy flux per unit mass, $Q_j/(\eta \pi r_j^2)\propto \Gamma_j^2$, is smaller, but the momentum flux, $\dot{M}j/(\pi r_j^2)\propto \eta \Gamma_j^2 \propto \eta_r$, is larger (see Equations (\ref{Gammaj})-(\ref{etar})). Consequently, in Q45-$\eta3$-r100, the Lorentz factor is smaller, $\Gamma_j = 1.9$, but the jet-head speed is higher, $v_{\rm head}^*/c \approx 0.048$, compared to $\Gamma_j = 11.2$ and $v_{\rm head}^*/c \approx 0.038$ in Q45-$\eta5$-r100. 
Thus, the Q45-$\eta3$-r100 model with a larger $\eta_r$ (and therefore a higher $v_{\rm head}^*$) produces a faster-moving jet head, resulting in a more elongated cocoon, despite the smaller average Lorentz factor of the spine, $\langle \Gamma \rangle_{\rm{spine}} \approx 1.8$.
This demonstrates that the two key parameters governing the morphology and flow dynamics of relativistic jets are $Q_j/\pi r_j^2$ and $\eta$ (or $\eta_r\propto \dot{M}_j/\pi r_j^2\propto \eta \Gamma_j^2$). 
These findings are consistent with previous numerical studies \citep[e.g.][]{rossi2008,perucho2017b,hardcastle2018,mukherjee2020}.} 

\subsubsection{ Evolution of Lobe Shape}

{Figure \ref{fig5} illustrates the time evolution of the actual jet-head speed, $v_{\rm head}$, the axial ratio $\mathcal{A}=l_{\rm jet}/\mathcal{W}$, and the lobe's lateral width $\mathcal{W}$ in our jet models. 
In the fiducial r10 models, $v_{\rm head}$ is smaller than $v_{\rm head}^*$, with $v_{\rm head}/v_{\rm head}^*\sim 0.5-0.7$ at $t_{\rm end}$ (see Table \ref{tab:t1}).  
In contrast, in models where the jet propagates into backgrounds of decreasing density and pressure, such as Q42-$\eta5$-r10-P and the three r100 models, $v_{\rm head}/v_{\rm head}^*\sim 1.5-3.0$ at $t_{\rm end}$. 
In the r100 models, $v_{\rm head}(t)$ increases after the jet-head crosses the core, as shown with the dark green, light blue, and dark blue lines in Figure \ref{fig5}(a). 
\citet{perucho2019a} demonstrated in their 3D simulations that the propagation of jet-head can accelerate in backgrounds with decreasing pressure and density, and also can be influenced by helical perturbations at the jet injection.
In our jet models, we add a small amplitude precession to the jet inflow in all cases. However, the r10 models do not show jet-head acceleration, suggesting that the impact of precession in the models is likely marginal.}

In Figure \ref{fig5}, we observe the following points:
(1) Among the fiducial r10 models, $v_{\rm head}$ is larger with higher $\eta_r\propto \eta \Gamma_j^2$.
(2) Among the r10 models, the axial ratio $\mathcal{A}$ is larger with higher $Q_j$.
(3) The differences in the Q42-$\eta5$-r10 and Q42-$\eta5$-r10-P models with the same $\eta_r$ indicate that the deceleration of the jet flow is less pronounced in a background with {faster} decreasing density/pressure.
(4) The differences in the Q45-$\eta5$-r100 and Q45-$\eta3$-r100 models demonstrate that a heavier jet with a larger $\eta_r$ propagates faster and generates a more elongated cocoon.

\begin{figure*}[t]
\centering
\includegraphics[width=0.95\linewidth]{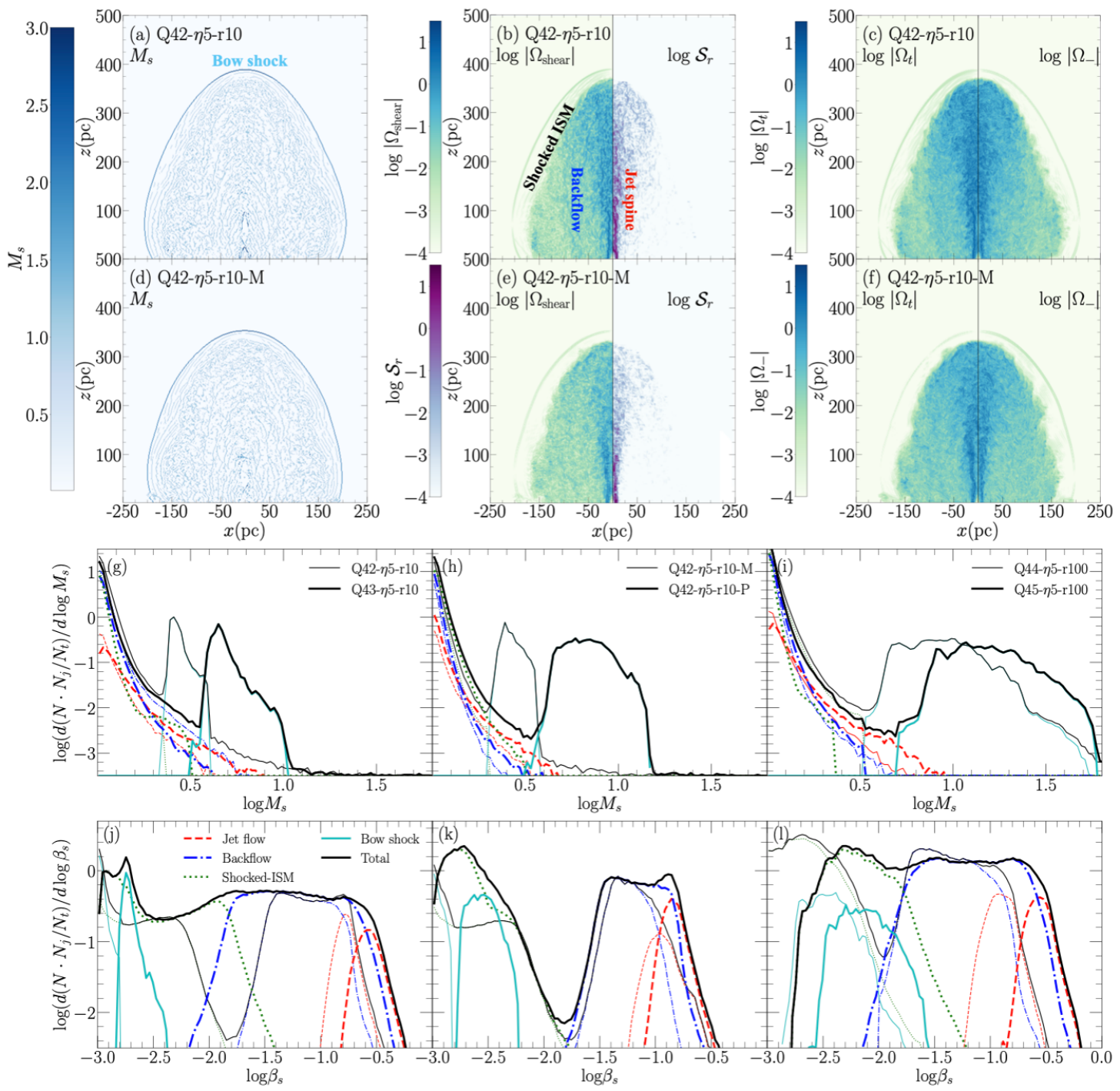}
\caption{
2D slice distributions of the Mach number ($M_s$) of shocks (a,d),
the magnitude of the velocity shear $\Omega_{\rm shear}$ (the left side with $x<0$ on panels (b) and (e)), the relativistic shear coefficient $\mathcal{S}_{r}$ (the right side with $x>0$ on panels (b) and (e)),
the magnitudes of the total vorticity $\Omega_t$ (the left side with $x<0$ on panels (c) and (f)), and the vorticity excluding the shear $\Omega_-$ (the right side with $x>0$ on panels (c) and (f)) for the Q42-$\eta5$-r10 and Q42-$\eta5$-r10-M models.
(g-i) PDFs of the shock Mach number, $M_s$, for the six models listed in Table \ref{tab:t1}.
(j-l) PDFs of the shock speed, $\beta_{s}= v_{s}/c$, for the same six models.
Different line types are used for the shock zones in the jet-spine flow (red dashed lines), backflow (blue dot-dashed lines), and shocked ISM (green dotted lines), and for the bow shock surface (cyan solid lines).
The black solid lines plot the PDFs for all the shocks identified in the simulation domain. 
The shock zones with $M_s\geq1.01$ are included, and the quantities shown are at $t_{\rm end}$. Here, $N_t$ is the total number of grid zones in the volume encompassed by the bow shock surface, and $N_j$ is the number of zones occupied by the initial jet radius.}
\label{fig6}
\end{figure*}

\subsection{Shocks, Shear, and Turbulence}\label{s3.3}

\begin{figure*}[t]
\centering
\includegraphics[width=0.9\linewidth]{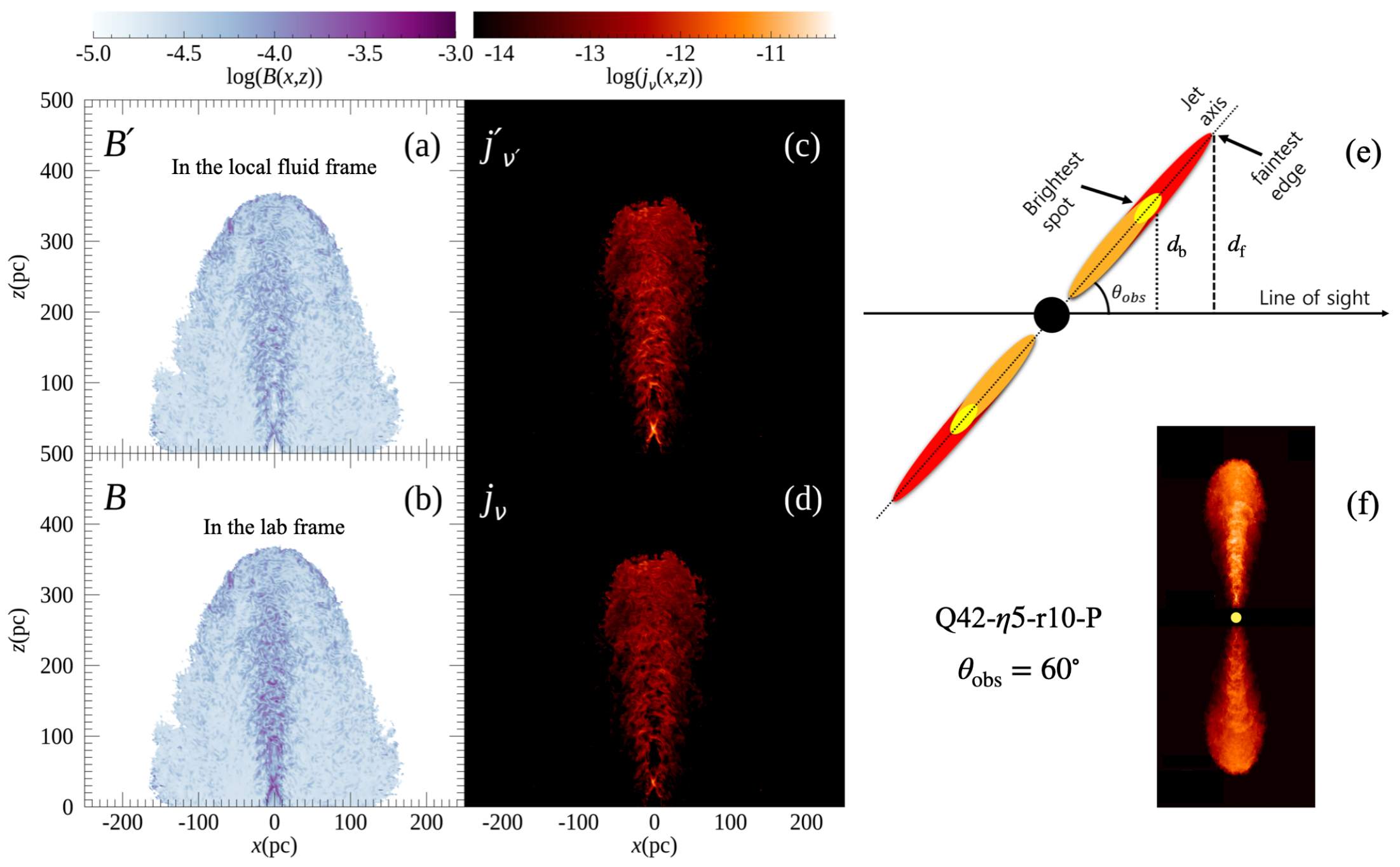}
\caption{
(a-b) 2D slice images of the magnetic field strength in the local fluid frame, $B^{\prime}(x,z)$, and in the computational frame, $B(x,z)$, at $t_{\rm end}$ for the Q42-$\eta5$-r10 model.
(c-d) 2D slice images of the synchrotron volume emissivity in the local fluid frame, $j^{\prime}_{\nu^{\prime}}(x,z)$ at $\nu^{\prime}=150$~MHz, and in the computational frame, $j_{\nu}(x,z)$ at $\nu=150$~MHz, for the same model.
 The images in (a-d) are in the $x-z$ plane through $y=0$.
(e) A schematic diagram of a pair of approaching and receding jets, tilted with the observation angle, $\theta_{\rm obs}$. The projected distances to the brightest spot, $d_b$, and to the faintest edge, $d_f$, from the injection point are illustrated. 
(f) Synchrotron surface brightness map, the log of the intensity calculated by integrating the synchrotron emissivity along the LOS, at $\nu=150$~MHz for a pair of approaching and receding jets displayed with $\theta_{\rm obs}=60^{\circ}$ for the Q42-$\eta5$-r10-P model. The unit is arbitrary. For illustrative purposes, a circle is inserted in the middle to represent the core region that contains the central black hole.
}
\label{fig7}
\end{figure*}

In Paper I, we investigated the properties of nonlinear flow structures, such as shocks, velocity shear, and turbulence, generated in the simulated FR-II jets. The primary objective of this analysis was to understand various particle acceleration processes operating in radio jets. Although the particle acceleration is not the main topic of this paper, we still present the following quantities utilizing the simulation data: shocks with their speed $\beta_s = u_s/c$ and Mach number $M_s$, velocity shear $\Omega_{\rm shear}\equiv |\partial v_z/\partial r|$, relativistic shear coefficient $\mathcal{S}_r=(\Gamma_z^4/15)\cdot \Omega_{\rm shear}^2$ where $\Gamma_z=[1-(v_z/c)^2]^{-1/2}$, vorticity $\Omega_t=|\mbox{\boldmath$\nabla$}\times\mbox{\boldmath$v$}|$, and vorticity excluding the shear $\Omega_{-} = |\mbox{\boldmath$\Omega_t$} + (\partial v_z/\partial r) \mbox{\boldmath$\hat{\theta}$}|$. For a comprehensive descriptions of these variables and the numerical procedures employed, readers are directed to the details provided in Paper I.

In Figures \ref{fig6}(a-f), we show the 2D distributions of shock zones, velocity shear parameters, and vorticity factors in the $x-z$ plane with $y=0$, for the Q42-$\eta5$-r10 and Q42-$\eta5$-r10-M models. Overall, the bow shock surface and recollimation shocks exhibit higher Mach numbers compared to the shocks formed in the backflow and the shocked background medium. In the models considered here, typically, the volume-averaged shock speeds are relativistic with $\langle \beta_s \rangle_{\rm JS}\approx 0.1-0.3$ in the jet spine, subrelativistic with $\langle \beta_s \rangle_{\rm BF}\approx 0.04-0.07$ in the backflow, and nonrelativistic with $\langle \beta_s \rangle_{\rm BS}\approx (1-7)\times 10^{-3}$ in the bow shock surface.

The velocity shear appears in the jet-spine and backflow. Especially, the relativistic shear coefficient is strongly concentrated along the interface between the jet-spine flow and the backflow owing to the $\Gamma_z^2$ factor: $\langle \mathcal{S}_r/(c/r_j)^2 \rangle_{\rm JS} \approx 0.01 - 0.06$ in the jet-spine flow, whereas $\langle \mathcal{S}_r/(c/r_j)^2 \rangle_{\rm BF} \approx (3-5)\times 10^{-3}$ in the backflow. The vorticity factors $\Omega_t$ and $\Omega_{-}$ spread out inside the cocoon. The volume-averaged vorticity excluding the shear, which is a more direct measure of turbulence, ranges as $\langle \Omega_{-}/(c/r_j) \rangle_{\rm JS} \approx 0.5–1.1$ in the jet-spine flow, and $\langle \Omega_{-}/(c/r_j) \rangle_{\rm BF} \approx 0.04-0.17$ in the backflow.
The flow dynamics in Figures \ref{fig6}(a-f) are mostly consistent with those in Figures 8, 11, and 13 of Paper I, which depicted high-power FR-II jets.

The comparison of Q42-$\eta5$-r10 and Q42-$\eta5$-r10-M, in Figures \ref{fig6}(a-c) and (d-f), respectively, shows relatively minor differences, except that the jet head advances more slowly in the model with mass loading. This indicates that the amount of mass loading, $q_D$, adopted here does not substantially affect the flow dynamics in this specific example.  

Furthermore, the probability distribution functions (PDFs) of $M_s$  and $\beta_s$ are presented in Figure \ref{fig6} (g-i) and (j-l), respectively, for six selected jet models.
{{The bow shock, recollimation shocks, and turbulent shocks are generally stronger, with higher $M_s$ and larger $\beta_s$, when induced by a jet with a larger initial Lorentz factor, $\Gamma_j$. Panels (g) and (j) confirm such trend:}}
the volume averaged values are $\langle \beta_s \rangle_{\rm BS}\approx 10^{-3}$, $\langle M_s \rangle_{\rm BS}\approx 2.7$, $\langle \beta_s \rangle_{\rm JS}\approx 0.17$, and $\langle M_s \rangle_{\rm JS}\approx 1.1$ in Q42-$\eta5$-r10, while $\langle \beta_s \rangle_{\rm BS}\approx 2\times 10^{-3}$, $\langle M_s \rangle_{\rm BS}\approx 4.6$, $\langle \beta_s \rangle_{\rm JS}\approx 0.27$, and $\langle M_s \rangle_{\rm JS}\approx 1.3$ in the higher-power Q43-$\eta5$-r10 model.
Panels (h) and (k) display the effects of mass loading and declining density/pressure background, respectively. On average, the shock speeds and Mach numbers are slightly smaller, with $\langle \beta_s \rangle_{\rm BS}\approx 10^{-3}$, $\langle M_s \rangle_{\rm BS}\approx 2.6$, $\langle \beta_s \rangle_{\rm JS}\approx 0.12$, and $\langle M_s \rangle_{\rm JS}\approx 1.1$ in the Q42-$\eta5$-r10-M model. On the other hand, they are substantially higher with $\langle \beta_s \rangle_{\rm BS}\approx 3\times 10^{-3}$, $\langle M_s \rangle_{\rm BS}\approx 6.7$, $\langle \beta_s \rangle_{\rm JS}\approx 0.15$, and $\langle M_s \rangle_{\rm JS}\approx 1.1$ in the Q42-$\eta5$-r10-P model.

The PDFs of $M_s$ and $\beta_s$ for the two r100 models on larger scales of $\sim 10$ kpc can be seen in Figures \ref{fig6}(i) and (l). 
{As expected, the higher-power jet, Q45-$\eta5$-r100, induces shocks with higher $M_s$ and larger $\beta_s$ on average, compared to the lower-power jet, Q44-$\eta5$-r100, as expected. 
In these two r100 models, the PDFs of $\beta_s$ and $M_s$ exhibit broader ranges than those in the fiducial r10 models shown in panels (g) and (j). {These broader distributions arise because the jets in the r100 models escape the dense core and propagate through the low-density regions of the stratified background, leading to shock formation in less dense environments. This results the PDFs that extend to higher $M_s$ and larger $\beta_s$.}}

\subsection{Radio Morphology of Jets}\label{s3.4}

\begin{figure*}[t]
\centering
\includegraphics[width=0.95\linewidth]{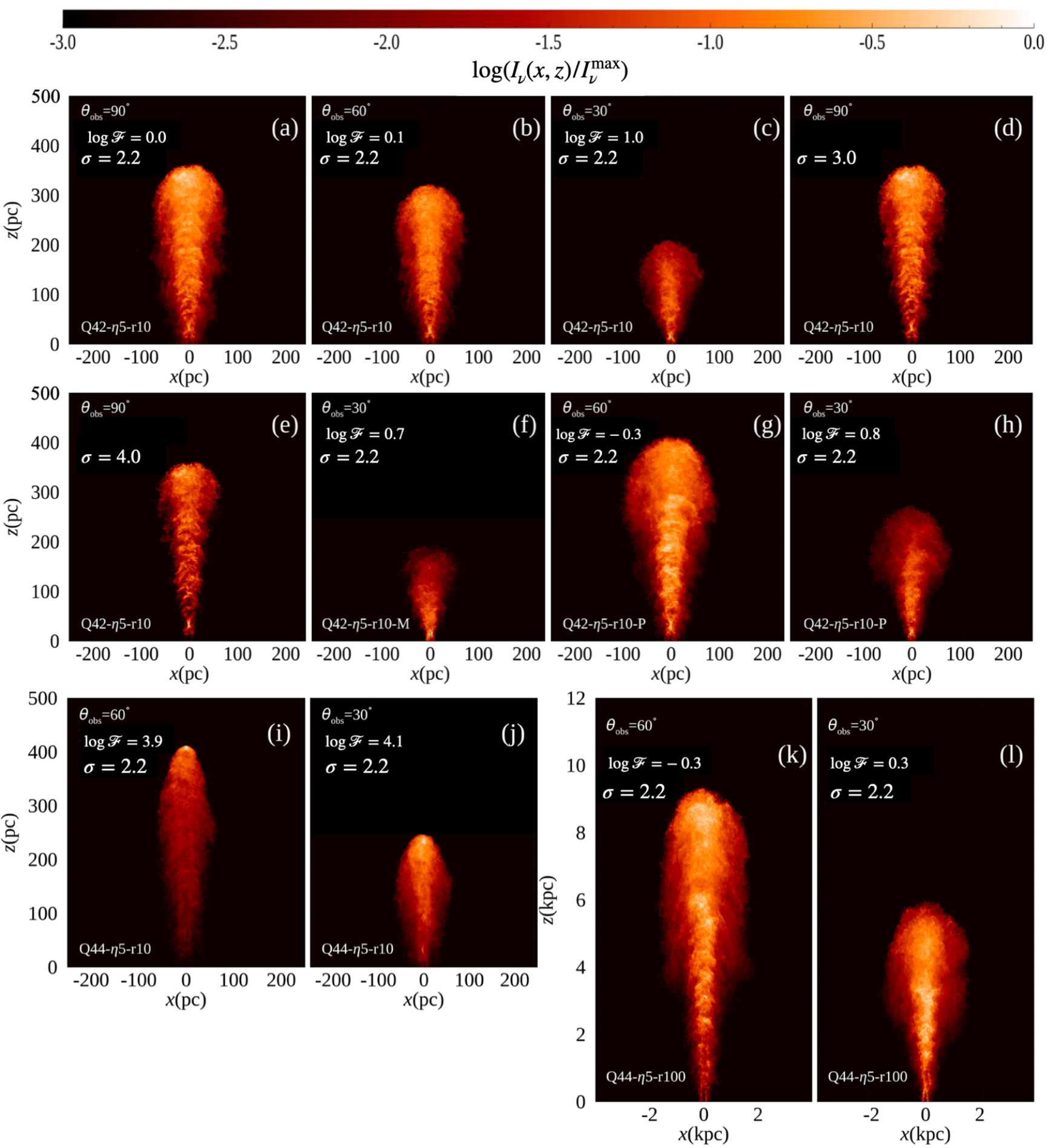}
\caption{2D maps of the synchrotron intensity integrated along the LOS, $\log[I_{\nu}(x,z)/I_{\nu}^{\rm max}]$, at $\nu_{\rm obs}=150~\rm MHz$, where
the intensity is normalized by its peak value, $I_{\nu}^{\rm max}$.
Each panel is labeled with the model name, $\sigma$, $\theta_{\rm obs}$, and the relative intensity ratio, $\mathcal{F}\equiv I_{\nu}^{\rm max}/I_{\nu,N}^{\rm Q42r10}$, where $I_{\nu,N}^{\rm Q42r10}$ is the peak value of the face-on map ($\theta_{\rm obs}=90^{\circ}$) for the Q42-$\eta5$-r10 model with $\sigma=2.2$. All maps are displayed at $t_{\rm end}$.
The mean Lorentz factor of the jet-spine flow, $\langle\Gamma\rangle_{\rm{spine}}$, is 2.1, 2.1, 1.7, 8.7, and 2.8 for Q42-$\eta5$-r10, Q42-$\eta5$-r10-P, Q42-$\eta5$-r10-M, Q44-$\eta5$-r10, and Q44-$\eta5$-r100, respectively (see Table \ref{tab:t1}).
Note that with relativistic beaming, the Doppler factor is $\mathcal{D}=1/(\Gamma[1-(v/c) \cos\theta])$.
So the emission is boosted most for mildly relativistic sources viewed at a small LOS angle, $\theta\lesssim 30^{\circ}$.
}
\label{fig8}
\end{figure*}

\begin{figure*}[t]
\centering
\includegraphics[width=0.95\linewidth]{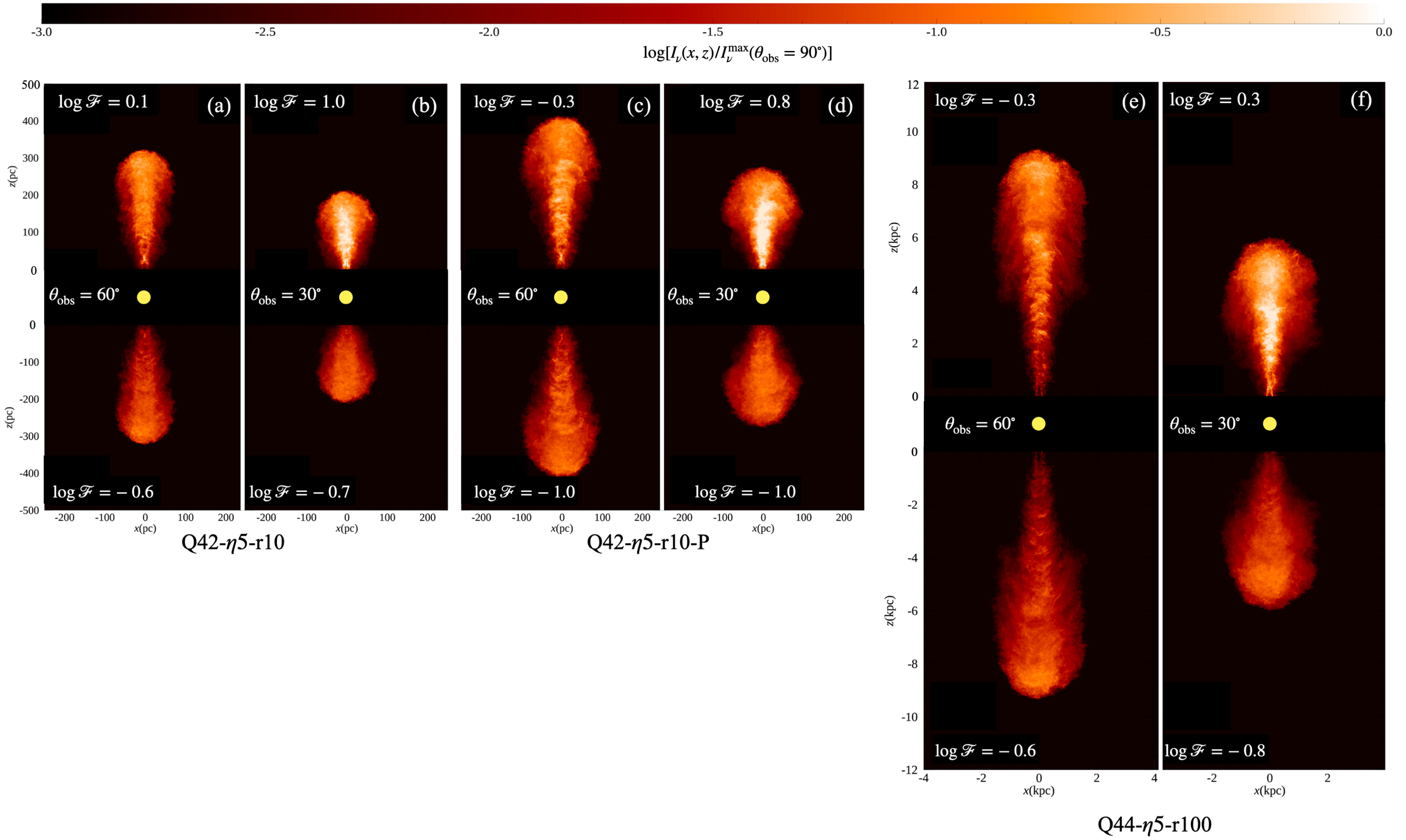}
\caption{2D maps of the synchrotron intensity, $\log[I_{\nu}(x,z)/I_{\nu}^{\rm max}(\theta_{\rm obs}=90^{\circ})]$, at $\nu_{\rm obs}=150~\rm MHz$ for the approaching jet (top) and receding counterjet (bottom) at $t_{\rm end}$ for the Q42-$\eta5$-r10, Q42-$\eta5$-r10-P, and Q44-$\eta5$-r100 models.
The intensity is calculated with $\sigma=2.2$. 
For each model, two cases with $\theta_{\rm obs}=60^{\circ}$ and $30^{\circ}$ are presented.
Each panel shows a pair of jet and counterjet, with intensities normalized by the peak value for each model as viewed at $\theta_{\rm obs}=90^{\circ}$, $I_{\nu}^{\rm max}(\theta_{\rm obs}=90^{\circ})$.
For both the jet and counterjet maps in each panel, the relative intensity ratios, $\mathcal{F}= I_{\nu}^{\rm max}/I_{\nu,N}^{\rm Q42r10}$, are provided, where $I_{\nu}^{\rm max}$ is the peak intensity of each jet (or counterjet), and $I_{\nu,N}^{\rm Q42r10}$ is the peak value of the face-on map ($\theta_{\rm obs}=90^{\circ}$) for the Q42-$\eta5$-r10 model.
Here, $\langle\Gamma\rangle_{\rm{spine}}\approx$~2.1, 2.1, and 2.8 for Q42-$\eta5$-r10, Q42-$\eta5$-r10-P, and Q44-$\eta5$-r100, respectively.
For illustrative purposes, a circle is inserted in the middle to represent the core region that contains the central black hole.}
\label{fig9}
\end{figure*}

The synchrotron surface brightness, or intensity, depends on factors such as magnetic fields, CR electron distributions, relativistic beaming, and relativistic Doppler shift. It determines the morphological characteristics of optically thin, diffuse sources like FR-type radio galaxies. {We employ models for magnetic fields and CR electrons that use hydrodynamic variables from simulations}, and estimate the synchrotron volume emissivity, $j_{\nu}(x,y,z)$, from which the synthetic intensity map, $I_{\nu}(x,z)$, is generated, as described in Section \ref{s2.4}.  
{We point out that our synthetic maps may not fully capture realistic radio morphology due to limitations in our modeling, such as the omission of MHD effects and the use of simplified methods for estimating magnetic field strength and CR electron populations.
Yet, as discussed in Paper I, our RHD simulations show that even high-power FR-II jets with $Q_j\approx 3\times 10^{47} {\rm erg~s^{-1}}$ form chaotic, turbulent structures near the jet head, rather than a well-defined termination shock (hot spot), acting as a stable working surface (see Figure 2 in that paper). 
None of the synthetic maps of the low-power models considered here exhibit distinct hot spots.
Despite these limitations, Figures \ref{fig7}-\ref{fig9} illustrate the general dependence of radio morphology on various model parameters, as described in detail below.} 

Figures \ref{fig7}(a-b) display the 2D distributions of magnetic field strength in two different frames: $B^{\prime}(x,z)$ in the local fluid frame and $B(x,z)$ in the computational (or observer) frame. The results at $t_{\rm end}$ are shown for the Q42-$\eta5$-r10 model. With our magnetic field recipes, both $B^{'}$ and $B$ are a few $\times~10~\mu$G in the subrelativistic backflow, while $B^{\prime}\sim ~10 \sim 100~\mu$G and $B\gtrsim100~\mu$G in the relativistic jet-spine flow. These are in good agreement with the magnetic field strength inferred from X-ray and radio observations of radio galaxies \citep[see, e.g.,][]{kataoka2005,ito2021}, {as well as the values extrapolated from the studies of objects on more compact scales \citep{zamaninasab2014}.}

Figures \ref{fig7}(c-d) depict the 2D slice views of the synchrotron volume emissivity, $j^{\prime}_{\nu^{\prime}}(x,z)$ in the fluid frame and $j_{\nu}(x,z)$ in the observer frame, respectively. The regions near the first and second recollimation shocks, along with those around the expanding boundary/shear layer, appear brighter, whereas the jet spine appears to be hollow. When viewed from the computational frame ($j_{\nu}$), faster-moving regions on the jet spine become fainter due to the Doppler dimming effect; hence, only the shear layer and the region near the jet head dominate the image. These features fit well with the expanding ``spine/shear-layer model'' proposed by \citet{laing2002a}. In this model, Doppler dimming, shock-mediated brightening, and strong emission from the shear/turbulent region can be seen. 

The surface brightness map is a superposition of such slices, integrated along the LOS. As an illustration, a radio map for Q42-$\eta5$-r10-P is displayed in Figure \ref{fig7}(f). Here, the simulation box is tilted by the angle of $\pi/2-\theta_{\rm obs}$ around the $x$-axis with the inclination angle $\theta_{\rm obs}=60^{\circ}$, as demonstrated in Figure \ref{fig7}(e). This example illustrates the pair of the boosted approaching jet and the deboosted receding jet. 

Figure \ref{fig8} shows how the morphology of radio jets at 150 MHz varies with the inclination angle, $\theta_{\rm obs}$, and the power-law index, $\sigma$, of the CR electron spectrum in the Q42-$\eta5$-r10, Q42-$\eta5$-r10-M, Q42-$\eta5$-r10-P, Q44-$\eta5$-r10, and Q44-$\eta5$-r100 models.
We note that with our modeling of synchrotron emission, only the relative intensity is meaningful in the surface brightness maps; thus, the intensity is normalized by its peak value, $I_{\nu}^{\rm max}$, to highlight the characteristic features in each panel. 
The relative intensity ratio, $\mathcal{F}= I_{\nu}^{\rm max}/I_{\nu,N}^{\rm Q42r10}$, is provided in each panel, where $I_{\nu,N}^{\rm Q42r10}$ represents the peak value of the face-on map ($\theta_{\rm obs}=90^{\circ}$) for the Q42-$\eta5$-r10 model with $\sigma=2.2$. 
For instance, $\mathcal{F}$ is higher for greater $Q_j$ and for smaller $\theta_{\rm obs}$.
In addition, the values of $\mathcal{F}|_{\theta_{\rm obs}=90^{\circ}}$ are listed in the fifth column of Table \ref{tab:t2}.

Figure \ref{fig8}(a) exhibits an ``edge-brightened'' morphology with a bright but diffuse jet head in the face-on case ($\theta_{\rm obs}=90^{\circ}$). Additionally, mixing layers around the jet spine are clearly visible. In contrast, Figure \ref{fig8}(c) shows a ``center-brightened'' morphology with a fainter diffuse lobe for $\theta_{\rm obs}=30^{\circ}$ due to strong relativistic beaming effects. 
Note that $\mathcal{F}=10$ in panel (c), so the region of the recollimation shocks is much brighter than that in panel (a).
As expected, the morphology at $\theta_{\rm obs}=60^{\circ}$ is intermediate, with both the center and edge displaying comparable brightness.

{ 
Following the original criterion established by \citet{fanaroff1974},} we estimate the FR ratio, $d_b/d_f$, by measuring the projected distances from the injection point to the brightest spot, $d_b$, and to the faintest edge, $d_f$, as defined in Figure \ref{fig7}(e).
We point out that the first recollimation shock is excluded in the estimation of $d_b$.
The comparison of Figure \ref{fig8} (a-c) demonstrates the transition from {edge-brightened to center-brightened} as $\theta_{\rm obs}$ decreases from $90^{\circ}$ to $30^{\circ}$, resulting in a decrease in the $d_b/d_f$ ratio from 0.94 to 0.24.
The FR ratios for inclination angles, $\theta_{\rm obs}=90^{\circ},~60^{\circ},~{\rm and}~30^{\circ}$, are given in Columns 6-8 of Table \ref{tab:t2}, respectively.  
For Q44-$\eta5$-r10, which displays {a hotspot at the jet head}, $d_b\approx d_f$, so we set $d_b/d_f\sim 1$ (see Figure \ref{fig8}(i-j)).

On the other hand, the comparison of Figures \ref{fig8}(a), (d), and (e) shows that, as the power-law slope $\sigma$ increases, the intensity map becomes progressively dominated by the jet material, characterized by higher $f_{\rm jet}$, greater $\epsilon$, and lower $\rho$. This is consistent with the scaling of the normalization factor in our model, $\mathcal{N}_0^{'}\propto f_{\rm jet}\cdot \epsilon (\epsilon/\rho)^{\sigma-2}$.
{We note that {for radio maps different $\sigma$'s}, the normalization of the synchrotron emissivity depends somewhat arbitrary on how we model the CR electron population, ${N}^{\prime}_{e}(\gamma_e^{\prime})$. Therefore, the relative intensity ratio $\mathcal{F}$ is not explicitly provided in Figures \ref{fig8}(d)-(e) for $\sigma=3$ and 4.}

The comparison between Figures \ref{fig8}(c) and (f) reveals the effects of mass-loading in Q42-$\eta5$-r10-M. In panel (f), the absence of a visible arch {at the jet head} makes the morphology akin to a tailed FR-I source \citep[e.g][Figure 2]{mingo2019}. 
In Figures \ref{fig8} (g-h), the morphology of the Q42-$\eta5$-r10-P jet with $\theta_{\rm obs}=60^{\circ}$ and $30^{\circ}$ is seen to be dominated by loss of pressure confinement, resulting in a bulge caused by what seems to be like a flaring activity. 

To analyze the asymmetry in the radio morphology between the approaching jet and the receding counterjet, we generate a counterjet in the region $z<0$ by mirroring the simulated jet in $z>0$, that is, by reversing the vertical velocity component, $v_z \rightarrow -v_z$.
Both the jet and the counterjet are tilted at different inclination angles, and surface brightness maps are generated, as illustrated in Figure \ref{fig7}(e). In this configuration, the approaching jet brightens while the receding counterjet dims, due to the relativistic bulk motion of the jet-spine flow.

In Figure \ref{fig9}, we present the surface brightness maps of jet-counterjet pairs at two different inclination angles for the Q42-$\eta5$-r10, Q42-$\eta5$-r10-P, and Q44-$\eta5$-r100 models.
The maps clearly show an asymmetry between the jet-counterjet pairs, characterized by the boosted approaching jet and the de-boosted receding jet. This asymmetry becomes more pronounced as $\theta_{\rm obs}$ decreases, leading to stronger center-brightening in the approaching jet and more noticeable edge-brightening in the receding counterjet.  These results align well with the established correlation between core dominance and inclination angle in 3CRR sources  \citep[e.g.,][]{marin2016}.

As shown in Figures \ref{fig9}(b), (d) and (f), at $\theta_{\rm obs}=30^{\circ}$ the receding counterjets may look like {edge-brightened jets}, whereas the approaching jets behave as BL Lac-like FR-I jets.
{For instance, Figure \ref{fig9}(f) resembles the radio image of the low-luminosity radio galaxy 3C 31 \citep{laing2002a}.}
Moreover, such pairs are similar to those seen in HYbrid MOrphology Radio Sources (HyMoRS) reported by \citet{gopalkrishna2000}. Recent observations by \citet{harwood2020} also found such boosted FR-I/de-boosted FR-II type hybrid sources, as we see in our synthetic maps. It is thought that the formation of HyMoRS is controlled by complicated processes, possibly involving the central engine and the properties of the host galaxy \citep{gopalkrishna2000}, apart from the inclination angle. However, our results indicate that the jet inclination angle may play a significant role in the FR-I/II hybrid sources.

\section{Summary}\label{s4}

The flow dynamics and radio morphology of low-power FR-I jets are primarily influenced by the deceleration and expansion of the jet-induced flow, resulting from the interplay between the jet propagation and interactions with the surrounding medium.

As discussed in Section \ref{s2.3}, these are governed by the bulk Lorentz factor, $\Gamma_j^2 \propto Q_j/\eta\pi r_j^2$, and the jet-head advance speed, $v_{\rm head}^*/c \propto \eta_r \propto \dot{M}_j/\pi r_j^2$, in the initial stage of jet evolution.
As the jet evolves, the jet-spine flow expands and decelerates, causing gradual decreases in both the mean Lorentz factor of the jet spine, $\langle\Gamma\rangle_{\rm{spine}}$, and the actual jet-head advance speed, $v_{\rm head}$.    
Moreover, the evolution of the jet-induced flows is influenced by the distribution of the background density and pressure, as both the entrainment of the ambient gas and the external density/pressure gradient affect the deceleration of the jet-spine flow. Notably, once the jet extends beyond the dense core of the host galaxy into the stratified halo with declining density and pressure, it undergoes a phenomenon known as flaring or spreading of the jet \citep[e.g.,][]{laing2002a}. 

Additionally, asymmetry in the radio morphology of jet-counterjet pairs is observed in certain FR-I radio galaxies; for instance, inner jets are often one-sided on small scales, while diffuse lobes appear on both sides on large scales \citep[e.g.,][]{urry1991,laing2014}. This asymmetry is {largely} attributed to relativistic Doppler beaming of the jet-spin flow, observed at inclination angles, $\theta_{\text{obs}}<90^{\circ}$. 

Understanding these various effects is crucial for comprehending the flow dynamics of relativistic jets and ultimately elucidating the FR-I/II dichotomy.
Using the high-order-accurate HOW-RHD code \citep{seo2021a}, we conducted 3D RHD simulations of low-power relativistic jets propagating through a galactic core surrounded by a stratified halo. 

We investigated the three groups of jet models, which are summarized in Table \ref{tab:t1}. The fiducial r10 models in the first group feature a jet-injection radius of 10 pc and the jet remaining within the galactic core. The two comparison r10 models are included in the second group.
The Q42-$\eta5$-r10-M model is designed to examine the impact of mass-loading due to stellar winds, while the Q42-$\eta5$-r10-P model represents the halo with a power-law ($\propto r^{-3/2}$) declining density and pressure.
The r100 models in the third group have a jet-injection radius of 100 pc and the jet propagating into the stratified halo.
A wide range of the jet parameters are considered: $Q_j=2.2\times 10^{42} - 3.5\times 10^{45}{\rm erg~s^{-1}}$ and $\eta=10^{-5}-10^{-3}$, resulting in $\langle\Gamma\rangle_{\rm{spine}}\approx 1.7 - 8.7$ and $v_{\rm head}/c\approx 0.005 - 0.15$.

As in Paper I, we examined the dynamical properties of nonlinear structures such as shocks, velocity shear, and turbulence.
Additionally, by employing models for magnetic field distribution and CR electron population, we estimated the synchrotron emission in the jet models. We then produced the radio surface brightness maps, $I_{\nu}(x,z)$, of the simulated jets as observed from various inclination angles, $\theta_{\rm obs}$, and explored the morphological properties of the maps.

The main findings are summarized below:

1. We confirm that the overall dynamics, structure, and morphology of relativistic jets are primarily governed by the initial Lorentz factor, $\Gamma_j$, and the jet-head advance speed, $v_{\rm head}^*$.
Alternatively, the evolution of jets in models where $p_j=p_c$ is controlled by three traditional jet parameters: $Q_j$, $r_j$, and $\eta$.
Low-power jets with small $\Gamma_j$ tend to undergo substantial deceleration and decollimation, facilitated by shock formation, jet expansion, and turbulent mixing. The deceleration factor, $\mathcal{R}_{\rm dec}\equiv v_{\rm head}/v_j$, is smaller for jets with lower $\Gamma_j$ (see Figure \ref{fig4}).

2. For the jet models with lower $v_{\rm head}^*$, the jet head advances more slowly and decelerates more strongly through the entrainment of ambient gas within the mixing layers induced by various instabilities. Additional mass loading from stellar winds could contribute to early deceleration of the jet within the galactic core, as demonstrated in the Q42-$\eta5$-r10-M model. 
However, with the mass-loading rate typically found in elliptical galaxies, the dynamical consequences on deceleration and decollimation are only marginal.

3. As the jet head moves into the stratified halo with decreasing density and pressure, the expansion in both the transverse and longitudinal directions causes the flaring of the jet flow, as demonstrated in Q42-$\eta5$-r10-P and Q44-$\eta5$-r100.
Our 3D RHD simulations capture the formation of the spine/shear layer that successfully describes the observational properties of FR-I jets \citep{laing2002a}. 

4. The radio morphology of the simulated jets depends on the inclination angle $\theta_{\rm obs}$. FR-I jets with mildly relativistic $\langle\Gamma\rangle_{\rm{spine}}$ and small $\theta_{\rm obs}$ could be highly boosted, resembling BL Lac objects \citep{urry1991}. 
Furthermore, jet-counterjet asymmetry is well captured in our synthetic radio maps, including the Doppler dimming of the receding counterjet, which matches very well with observed double-lobed FR-I jets \citep{laing1999,laing2014}. Our results also confirm the idea that the hybrid morphology (FR-I/II) can be found in FR-I sources with small $\theta_{\rm obs}$ \citep{gopalkrishna2000}.

\begin{acknowledgments}
{The authors would like to thank the anonymous referee for constructive comments and suggestions.}
This work was supported by the National Research Foundation (NRF) of Korea through grants 2020R1A2C2102800, 2022R1I1A1A01065435, 2023R1A2C1003131, and RS-2022-00197685.
\end{acknowledgments}

\bibliography{FR-I_References}{}

\bibliographystyle{aasjournal}

\end{document}